\documentclass[a4paper]{article}

\usepackage{cite}				

\usepackage[utf8]{inputenc}
\usepackage[T3,T1]{fontenc}
\usepackage[french,english]{babel}
\usepackage[noenc]{tipa}
\usepackage{tipx}
\usepackage[geometry,weather,misc,clock]{ifsym}
\usepackage{pifont}
\usepackage{eurosym}
\usepackage{amsmath,bm}
\usepackage{wasysym}
\usepackage{amssymb,amsmath,amsfonts,textcomp}
\usepackage{mathrsfs}
\usepackage{enumitem}

\DeclareMathAlphabet\mathbfcal{OMS}{cmsy}{b}{n}

\DeclareBoldMathCommand\boldlangle{\left\langle} 
\DeclareBoldMathCommand\boldrangle{\right\rangle} 

\usepackage[subrefformat=parens,labelformat=parens]{subfig}

\usepackage{float}

\usepackage{color}
\usepackage{array}
\usepackage{hhline}

\usepackage{subfig}

\usepackage[usenames,dvipsnames,svgnames,table]{xcolor}
\usepackage{soul}
\usepackage{pdfcomment}
\usepackage[mathlines]{lineno}

\usepackage{verbatim} 
 
\newcommand{\chapeau}[1]{\skew{3}\widehat{#1}}

\usepackage[pdftex]{graphicx}

\usepackage{calc}
\makeatletter
\newcommand\ps@Standard{
  \renewcommand\@oddhead{}
  \renewcommand\@evenhead{}
  \renewcommand\@oddfoot{\thepage{}}
  \renewcommand\@evenfoot{\@oddfoot}
  \renewcommand\thepage{\arabic{page}}
}
\makeatother
\newlength{\idxmathdepth}\newlength{\idxmathtotal}\newlength{\idxmathwidth}\newlength{\idxraiseme}

\newcommand\apos[1]{{#1}^{\prime \!}}

\newcommand\blue[1]{{\color{blue}#1}}

\newcommand\distrho{\mbox{\small\( \rho^\mu \)}}
\newcommand\distxi{\mbox{\small\( \xi^\mu \)}}
\newcommand\distchi{\mbox{\small\( \chi \)}}
   
\usepackage[normalem]{ulem}
\newcommand\trait{\bgroup\markoverwith{\textcolor{red}{\rule[0.5ex]{2pt}{0.75pt}}}\ULon}

\usepackage{cancel}

\makeatletter
\def\uwave{\bgroup \markoverwith{\lower3.5\p@\hbox{\sixly \textcolor{red}{\char58}}}\ULon}
\font\sixly=lasyb10 scaled 652
\makeatother

\newcommand{\bref}[1]{\blue{\fontsize{10}{10}\selectfont{\ref{#1}}}} 

\let\origbref\bref
\renewcommand{\bref}[1]{\unskip~\origbref{#1}} 				

\renewcommand*{\eqref}[1]{{\color{blue}\!\!~(\ref{#1})}}                    

\usepackage{cleveref}
\crefrangeformat{equation}{(#3#1#4)--(#5#2#6)}
\crefformat{equation}{(#2#1#3)}

\date{}

\begin{document}

\title{\textbf{Mean field analysis of large-scale interacting populations of stochastic conductance-based spiking neurons using the Klimontovich method}}
\author{Daniel Gandolfo, Roger Rodriguez\\  [3pt]
{\small Aix Marseille Univ, Université de Toulon, CNRS, CPT, Marseille, France} \\ [3pt]
{\small   gandolfo@cpt.univ-mrs.fr,\  rodrig@cpt.univ-mrs.fr}\\ \\
Henry C. Tuckwell{\;$^{\tiny 1, 2}$}\\  [3pt]
{\small $^1$ School of Electrical and Electronic Engineering, University of Adelaide,}\\
{\small Adelaide, South Australia 5005, Australia} \\ [3pt]
{\small $^2$ School of Mathematical Sciences, Monash University,}\\ 
{\small Clayton, Victoria 3168, Australia}\\ [3pt]
{\small  Henry.Tuckwell@Adelaide.edu.au}}

\maketitle

\begin{abstract}
We investigate the dynamics of large-scale interacting neural populations, composed of conductance based, spiking model neurons with modifiable synaptic connection strengths, which are possibly also subjected to external noisy currents. The network dynamics is controlled by a set of neural population probability distributions ($\mathrm{PPD}$) which are constructed along the same lines as in the Klimontovich approach to the kinetic theory of plasmas. An exact non-closed, nonlinear, system of integro-partial differential equations  is derived for the $\mathrm{PPD}$s.
As is customary,  a closing procedure leads to a mean field limit. The equations we have  obtained are of the same type as those which have been recently derived using rigorous techniques of probability theory. 
The numerical solutions of these so called McKean-Vlasov-Fokker-Planck equations, which are only valid  in the limit of infinite size networks, actually shows that the statistical measures as obtained from $\mathrm{PPD}$s are in good agreement with those obtained through direct integration of the stochastic dynamical system for large but finite size networks.
Although numerical solutions have been obtained for networks of Fitzhugh-Nagumo model neurons, which are often used to approximate Hodgkin-Huxley model neurons, the theory can be readily applied to networks of general conductance-based model neurons of arbitrary dimension.	  
 \end{abstract}
{\bf  Classifications (2016).}

{\bf{Key words.} Computational neuroscience.  
Conductance-based neural models. Fitzhugh-Nagumo model. Stochastic differential equations. Klimontovich method. Mean-field limits. Neural networks. 
}

\section {Introduction}
\label{sec:intro}
		A quantitative understanding of brain neuronal activity 
   requires the construction of detailed models of neuronal networks, the development and application of powerful numerical tools and mathematical techniques to analyze them.
The large number of cells and their complex anatomy and physiology  present  challenges which must be met  in order to achieve these goals. Thus even though  there is much experimental detail available on the electrophysiology and neurochemistry of individual cells, incorporating these data at the macroscopic level is generally intractable.
In order to make some progress in this
endeavor, it is therefore useful to make several simplifications at the single cell and population levels. 
We find it useful to employ mathematical  methods from 
the kinetic theory of plasmas\blue{\cite{Klimon,Ichima,Nichol}}, whose concepts will be developed here within the framework of brain nervous activity. 
Such an approach was used recently for deterministic neural networks\blue{\cite{Buice2}}.

		With large populations of neurons the goal is a probabilistic description which predicts, not the behavior of any given neuron in a particular state, but is able to give sufficiently precise estimates of the probability that any individual neuron is in a particular state, or equivalently,  the proportion of neurons in a specified state.
The probabilistic approach of the network dynamics has a long history in computational neuroscience. It has been found to be essential in the analysis and description of many aspects of brain functioning\blue{\cite{Tuck1,Tuck2,Tuck3,Tuck4,Ermon,RodTuck1}}, at the levels of ion channel, single neuron and neuronal population.  Probabilistic descriptions  are also important in studies of the mechanisms of synchronization appearing in many fields of physics as shown by Kuramoto\blue{\cite{Kur1,Kur2,KurNish}}. Using this approach, neural populations, which can be viewed as weakly connected oscillators, have often been analyzed\blue{\cite{MirStrog,Abbott1,Daido,Buice1,Buice2}}. Other neural systems including integrate and fire neuron models, and their generalizations, have also  been described within probabilistic frameworks\blue{\cite{Tuck5,Tranch1,NiCamp1,NiCamp2,Knight,Tranch2,Hansel}}.

One of the main features of our approach is to incorporate, at the level of the individual neurons, a large degree of biological plausibility by describing them according to extended Hodgkin-Huxley (HH) frameworks\blue{\cite{HH,RodTuck2,TucRod}}. About connectivity between cells, our model incorporates neuroplasticity properties through the introduction of connection variables that are potential dependent. A stochastic dynamical system is built for such cells which are organized in different interacting populations. Then, the concept of set variables which is adapted from Klimontovich’s approach\blue{\cite{Klimon,Ichima,Nichol}} of the kinetic theory of plasmas is introduced.
Neural population probability distributions ($\mathrm{PPD}$) are defined as expectations of these set variables with respect to the probability distribution of the full interacting system which evolves through a Fokker Planck equation. 
A hierarchical set of non closed exact coupled equations for the different moments of the set variables can be derived. 
However, these Fokker-Planck type equations are rather difficult to analyze and numerically simulate.   
		To overcome this difficulty, our approach uses the rather traditional framework of mean field theory. If the number of cells in the network is made arbitrarily large, these cells being distributed in different populations, a set of non local, nonlinear integro-partial differential equations $\mathrm{(IPDE)}$ can be derived for the $\mathrm{PPD}$s.  
 It is interesting to observe that 
    the resulting equations we obtain, are of the same type as the McKean-Vlasov-Fokker-Planck (MVFP) equations which have been recently derived \blue{\cite{ToubFaug,Toub1,Toub2,FaugToubCessac}}
 using rigorous techniques of probability theory (see also\blue{\cite{DeMasi}}). 
In the case of Fitzhugh-Nagumo (FN) 
neural systems, we have computed the solution of the $\mathrm{PPD}$s in both cases of globally coupled and time dependent inhomogeneously connected networks.
		
In  section \bref{sec:popdynsyst}, we present the derivation of the $\mathrm{IPDE}$ suitable for the analysis of interacting populations of conductance-based spiking model neurons of the (HH) type. In section \bref{sec:applFN}, the special case of networks of  Fitzhugh-Nagumo model neurons is considered. 
Various statistical measures are defined which are suitable for the analysis of the dynamical behavior of these neuronal populations from a direct integration of the stochastic system and from the evaluation of solutions of $\mathrm{IPDE}$.
Excitatory and inhibitory uniformly connected networks are analyzed as well as excitatory non uniformly time dependent connected networks.
Moreover, the case of two large-scale interacting Fitzhugh-Nagumo populations is considered. 
Section \bref{sec:Num} describes the methods and parameters used in the simulations presented in section \bref{sec:applFN}. 
Conclusion follows in section \bref{sec:conclusion}.

\section{Integro-partial differential equations for population dynamical systems of coupled large-scale conductance-based neural networks.}
\label{sec:popdynsyst}
We consider networks of conductance-based model neurons with non-uniform synaptic connectivity. These networks are organized into a set of populations with interactions both within and between them. 
Let us call 
$\mathscr{P}_{\gamma}$,  ${\gamma=1,2,\ldots, P}$ a typical population, which is composed of $K_{\gamma }$ cells.

\subsection{The dynamical system for a given population  \texorpdfstring{$\mathscr{P}_{\gamma}$}{Pg}}
\label{subsec:DynSystOnePop}
The state variables of each cell ${i=1,2,\ldots, K_{\gamma }}$ in  $\mathscr{P}_{\gamma}$ are its membrane potential $V_i^{\gamma }$ and $\widehat{R}{_i^{\gamma }}$ which is an $m$-dimensional set of auxiliary variables.
 The dynamical laws for these auxiliary variables are of the Hodgkin Huxley type 
\begin {equation}
\frac {d\widehat{R}{_i^{\gamma }}}{dt}=\widehat{\Psi}{^{\gamma} }(V_i^{\gamma },\widehat{R}{_i^{\gamma }}).
\label{vars}
\end{equation}
More precisely, the $j^{th}$ component of the $m$--dimensional  vector $\widehat{R}{_i^{\gamma }}$ evolves according to laws of the form 
\begin{equation}
\frac {d}{dt}(\widehat{R}{_i^{\gamma }})_j=\widehat{\Psi}{^{\gamma} }(V_i^{\gamma },\widehat{R}{_i^{\gamma }})_j= A_j^1(V_i^{\gamma })(1-(\widehat{R}{_i^{\gamma }})_j)+A_j^2(V_i^{\gamma })(\widehat{R}{_i^{\gamma }})_j
\label{recovery}
\end{equation}
where $A_j^1, A_j^2:  \mathbb{R} \rightarrow  \mathbb{R}, \,j=1,2, \ldots, m$ are nonlinear functions of the membrane potential.

We call $\displaystyle {I}^{\gamma}(V_i^{\gamma },\widehat{R}{_i^{\gamma }})$ the set of ionic, active channels and passive leak, currents across the membrane of the $i^{th}$ cell. External currents $\mathcal {I}^{\gamma}_{i,ext}(t)$  may also be applied on all cells of the network. They are composed of a deterministic part $I^{\gamma}_{i,ext}(t)$ and a stochastic part $\eta _{i,t}^{\gamma}$, 
\begin{equation}
\mathcal {I}^{\gamma}_{i,ext}(t)=I^{\gamma}_{i,ext}(t)+\eta _{i,t}^{\gamma}
\end{equation}
where $\eta _{i,t}^{\gamma}$ is a white noise such that 
\begin{equation}
\langle\eta _{i,s}^{\gamma}\eta _{j,t}^{\gamma}\rangle\ =\ \delta_{ij}\, \delta(s-t)\, \beta^{\gamma}_{i,V}, \ \  i,j=1,2,\dots, K_{\gamma},
\label{noise}
\end{equation}
where $\delta_{ij}$ is the Kronecker symbol,  $\delta(\cdot)$ is the delta distribution and $\beta^{\gamma}_{i,V}  \ \  i=1,2,\dots, K_{\gamma}$ are noise parameters.
As in\blue{\cite{ToubFaug}}, we assume that the deterministic part of the current and the noise parameters are the same for all neurons in each given population, $I^{\gamma}_{i,ext}(t)=I^{\gamma}_{ext}(t)$ and $\beta^{\gamma}_{i,V}=\beta^{\gamma}_{V}$, $i=1,2,\dots, K_{\gamma}$.

Let us now introduce the connectivity properties of the model. Highly nonuniform connectivity has been observed in many biological neural networks. 
For example, in\blue{\cite{Song}}, using statistical techniques, structural features of synaptic connectivity have been shown to be quite different for local cortical circuitry from those of random networks.
In\blue{\cite{McLaughlin}}, a computational model has been developed for the primary visual cortex where emphasis is given to the role of cortical architecture,  particularly through synaptic connectivity, in the functional cortical activity.

Our approach to the variability of the connection strengths between cells is made in terms of new variables entering in the description of synaptic currents.
More precisely, let us call
$I^{\gamma}_{syn,i}(t)$  the current entering the $i^{th}$ cell as a result of the firing activity of all other cells in the network,  
\begin{equation}
I^{\gamma}_{syn,i}(t) =
\frac {1}{K_{\gamma}}\sum _{j=1}^{K_{\gamma }}M(V_i^{\gamma },\widehat{R}{_i^{\gamma }},\phi_i^{\gamma}, V_j^{\gamma },\widehat{R}{_j^{\gamma }},\phi_j^{\gamma})
\label{Isyn}
\end{equation}
where the bounded coupling functions $M$ will be given explicitly below.

The connection coefficients between pairs $(i,j)$ are scaled by the factor $\frac {1}{K_{\gamma}}$, so any divergence of this interaction term  is prevented when the network size increases  (see e.g.\blue{\cite{Kur1}},\blue{\cite{ToubFaug}}).
In our approach, the synaptic current  $I^{\gamma}_{syn,i}(t)$ depends also on real functions $\phi_i^{\gamma}(t)$ and $\phi_j^{\gamma}(t)$ which characterize the time dependent non homogeneous coupling  $J_{ij}^{\gamma}(t)$ between cells $i$ and $j$ through a suitable function  $\Gamma^{\gamma}$, namely 
 \begin{equation}
J_{ij}^{\gamma}(t) = \Gamma^{\gamma}(\phi_i^{\gamma}(t), \phi_j^{\gamma}(t)).
\label{Jij}
\end{equation}
Neurons in neocortical and other networks in the brain usually have specific sets of afferent inputs with a variety of strengths (see for example\blue{\cite{McLaughlin}}). 
 
Moreover, it is assumed  that there is a dynamical law for these $\phi_i$ variables of the form
\begin{equation}
\frac{d \phi_i^{\gamma}}{dt} = \Omega^{\gamma}(\phi_i^{\gamma},V_i^{\gamma},\widehat{R}{_i^{\gamma }})
\label{Omega}
\end{equation}
for a  suitable function $\Omega^{\gamma}$.  In what follows, the functions $\Gamma^{\gamma}$ and $\Omega^{\gamma}$  will be chosen in such a way to ensure positivity of couplings and  the bounded variation of $\phi$-variables.
In that way, the variables $\phi_i^{\gamma}$ (resp $\phi_j^{\gamma}$) which build the connection matrix  $J_{ij}^{\gamma}(t)$ as time evolves, depend on postsynaptic  (resp. presynaptic) activity. In the following, we shall consider homogeneous networks for which these $\phi$ variables are constant and inhomogeneous networks
for which are considered specific simple expressions for the functions $\Gamma^{\gamma}$ and $\Omega^{\gamma}$. In both cases, the forms of ${I}^{\gamma}$ and $I^{\gamma}_{syn,i}$ will be specified.

Finally, let us summarize the structure of the network  dynamical system (omitting the capacitance parameter). 
For $i=1,2,\dots,K_\gamma$, the system can be written
\begin{align}
\frac{dV_i^{\gamma}}{dt} &=I^{\gamma}(V_i^{\gamma },\widehat{R}{_i^{\gamma }})+I^{\gamma}_{ext}(t) + I^{\gamma}_{syn,i}(t) +\eta _{i,t}^{\gamma} \nonumber \\
\frac {d\widehat{R}{_i^{\gamma }}}{dt} &=\widehat{\Psi}{^{\gamma} }(V_i^{\gamma },\widehat{R}{_i^{\gamma }})   \label{syststochgenVect}  \\
\frac{d \phi_i^{\gamma}}{dt} &= \Omega^{\gamma}(\phi_i^{\gamma},V_i^{\gamma},\widehat{R}{_i^{\gamma }})  \nonumber \\
\intertext{which can be put in the  following vector form}
\frac{dZ_i^\gamma}{dt} &=\mathcal{F}^{\gamma}(Z_i^{\gamma })+\frac {1}{K_{\gamma}}\sum _{j=1}^{K_{\gamma }}\mathcal{M}^{\gamma}(Z_i^{\gamma},Z_j^{\gamma })+\zeta _{i,t}^{\gamma} \label{syststochgen} 
\end{align}

\noindent where the vectors $Z_i^\gamma, \; \mathcal{F}^{\gamma}(Z_i^{\gamma }), \; \zeta _{i,t}^{\gamma}, \; \mathcal{M}^{\gamma}(Z_i^{\gamma},Z_j^{\gamma })$  in $\mathbb{R}^{m+2}$ are given by
\begin{align}
Z_i^\gamma &=(V_i^\gamma,\widehat{R}{_i^{\gamma }},\phi_i^\gamma ) \label{Zigamma}\\
\mathcal{F}^{\gamma}(Z_i^{\gamma }) &=(I^{\gamma}(V_i^{\gamma },\widehat{R}_i^\gamma)+I_{ext}^\gamma(t),\; \widehat{\Psi}{^{\gamma} }(V_i^{\gamma },\; \widehat{R}{_i^{\gamma }}),\Omega^{\gamma}({\phi}{_i^{\gamma }},V_i^{\gamma },\widehat{R}{_i^{\gamma }})) \\
\mathcal{M}^{\gamma}(Z_i^{\gamma},Z_j^{\gamma }) &=(M^{\gamma}(V_i^{\gamma },\widehat{R}{_i^{\gamma }}, \phi_i^{\gamma}, V_j^{\gamma },\widehat{R}{_j^{\gamma }},\phi_j^{\gamma}),\mathit{\chapeau{0}},\mathit{0})  \label{Mgamma}\\
\zeta _{i,t}^{\gamma} &=(\eta _{i,t}^{\gamma},\mathit{\chapeau{0}},\mathit{0}) 
\end{align}
$\chapeau{\mathit{0}}$ being the null vector in $\mathbb{R}^m$. 

Note that in \eqref{Mgamma}, $M^{\gamma}$ includes the coupling term $\Gamma^{\gamma}(\phi_i^{\gamma}(t), \phi_j^{\gamma}(t))$ (see \eqref{Jij}) accounting for the amplitude of the synaptic conductance involved in the connection between the presynaptic cell $j$ and postsynaptic cell $i$.  

\medskip

The structure established in $\eqref{syststochgen}$ is now generalized to a set of coupled neural populations.

\subsection{The mean field approach  for coupled conductance-based neural populations}
\label{subsec:MFHH}
One considers non uniformly 
 connected HH 
 type neurons which are organized in a set of $P$ populations. The dynamical system governing the state of this set, at each time $t$, assumes the following form
\begin{equation}
\frac{dZ_i^{\gamma }}{dt}=\mathcal{F}^{\gamma}(Z_i^{\gamma })+\zeta _{i,t}^{\gamma}+\sum _{\alpha=1}^{P}\frac{1}{K_\alpha}\sum _{j=1}^{K_{\alpha }}\mathcal{ M^{\gamma \alpha}}(Z_i^{\gamma},Z_j^{\alpha }), \ \ i=1,2, \ldots, K_{\gamma}, \ \ \gamma=1,2,\ldots, P.
\label{genedynsystPOP}
\end{equation}
We now derive mean field population equations for $\eqref{genedynsystPOP}$.

Let us denote by  
\begin{equation}
p_t((Z_i^{\alpha})_{i,\alpha })=p_t((Z_i^1)_{i=1,2,\ldots, K_1},\ldots,(Z_j^P)_{j=1,2,\cdots, K_P}) 
\label{Z}
\end{equation}
the joint probability distribution of the stochastic variables  $(Z_i^{\alpha })_{i,\alpha }={(Z_i^{\alpha})}_{i=1,2,\ldots, K_{\alpha }}^{\alpha =1,2,\ldots, P}$  and introduce the notation 
$$
\frac{\partial }{\partial Z}(\mathcal{H}(Z))=\sum _{j=1}^{m+2}\frac{\partial }{{\partial Z}^j}(\mathcal{H}(Z)^j) 
$$
where $Z=(Z^j)_{j=1,2, \ldots, m+2} \in \mathbb{R}^{m+2}$ and $\mathcal{H}(Z)=(\mathcal{H}(Z)^j)_{j=1,2, \ldots, m+2} \in \mathbb{R}^{m+2}.$

The Fokker Planck equation for the system $\eqref{genedynsystPOP}$ is
\begin{gather}
\frac{\mathit{\partial }}{\mathit{\partial }t}p_t((Z_i^{\alpha })_{i,\alpha })=-\sum _{\gamma =1}^P\sum _{i=1}^{K_{\gamma }}\frac{\partial }{\partial Z_i^{\gamma }}(\mathcal{F}^{\gamma}(Z_i^{\gamma})\, p_t((Z_i^{\alpha})_{i,\alpha }))
\label{equa3} \\
-\sum _{\gamma =1}^P\sum _{i=1}^{K_{\gamma }}\frac{\partial }{\partial Z_i^{\gamma }}\biggl(\sum _{\alpha=1}^P\frac 1{K_{\alpha}}\sum _{j=1}^{K_{\alpha }}\mathcal {M}^{\gamma \alpha }(Z_i^{\gamma },Z_j^{\alpha })\, p_t((Z_i^{\alpha 
})_{i,\alpha })\biggr)+\frac 1 2\sum _{\gamma =1}^P\sum _{i=1}^{K_{\gamma }}\, (\beta _V^{\gamma})^2\frac{\partial ^2}{\partial (V_i^{\gamma})^ 2}p_t((Z_i^{\alpha})_{i,\alpha }) \nonumber
\end{gather}

We now adopt the Klimontovich approach\blue{\cite{Ichima}} which has been successfully developed for the kinetic theory of gases and plasmas. The method is here adapted to the derivation of a probabilistic description of the system $\eqref{genedynsystPOP}$ of noisy interacting spiking neural populations (see also\blue{\cite{Buice2}}). One defines  the following set variable 
\begin {equation}
\widehat{n}{^{\mu }}(U)=\frac 1{K_{\mu }}\sum _{i=1}^{K_{\mu }}\delta(Z_i^{\mu }-U)
\label{setvar}
\end{equation}   
where $\delta (\cdot)$ is the Dirac distribution and $Z_i^{\mu }$ is the solution of  $\eqref{genedynsystPOP}$  written for the population $\mathscr{P}_{\mu}$. $Z_i^\mu$ and $U$ (see \eqref{Zigamma}) are in $\mathbb{R}^{m+2}$. In what follows, the following notation is used: $U=(u,\hat z,s), u,s$ in $\mathbb{R}$, $\hat z$ in $\mathbb{R}^{m}$. The expectation value of the stochastic variables 
$\widehat{n}{^{\mu }}(U)$ with respect to the probability distribution  $p_t$ is denoted by 
$n^{\mu}(U,t)$, so that 
\begin {equation}
n^{\mu }(U,t)=\langle\widehat{n}{^{\mu }}(U)\rangle_{p_t}.
\label{expect}
\end{equation}

We call $n^{\mu }(U,t)$ the neural population probability distribution $\mathrm{(PPD)}$ for the population $\mathscr{P}_{\mu}$.
We now derive an equation  for $n^{\mu }(U,t)$. The time derivative  $\frac{\partial }{\partial t}n^{\mu }(U,t)$ is composed of 3 terms
\begin {equation}
\frac{\mathit{\partial }}{\mathit{\partial }t}n^{\mu }(U,t)=\gamma^\mu_1+\gamma^\mu _2+\gamma^\mu _3.
\label{equa5}
\end{equation}

Let us consider each of them separately. The first one, $\gamma^\mu _1$  is given by 
\begin {equation}
\gamma^\mu _1=-\int _{\mathit{{R^{m+2}} }}\prod _{\delta =1}^{P}\prod _{l=1}^{K_{\delta }}dZ^{\delta }_l\sum
_{\gamma =1}^P\sum _{i=1}^{K_{\gamma }}\frac{\partial }{\partial Z_i^{\gamma }}\left\{ \mathcal{F}^{\gamma }(Z^{\gamma}_i)p_t((Z_i^{\alpha })_{i,\alpha})\right\}\frac 1{K_{\mu }}\sum _{j=1}^{K_{\mu }}\delta (Z_j^{\mu }-U).
\label{equa5a}
\end{equation}

Being a probability distribution, $p_t$ has nice vanishing properties for sufficiently large values of the variables  $Z_i^{\mu }$. 
Thus, a simple integration by parts on \eqref{equa5a} enables us to deduce the following expression 
for  $\gamma^\mu _1$ 
\begin {equation}
\gamma^\mu _1=\int _{\mathit{{R^{m+2}} }}\prod _{\delta =1}^{P}\prod _{l=1}^{K_\delta }dZ^{\delta }_l\sum
_{\gamma =1}^P\sum _{i=1}^{K_{\gamma }}\left\{\mathcal{F}^{\gamma }(Z^{\gamma }_i)p_t((Z_i^{\alpha })_{i,\alpha})\right\}\frac{\partial }{\partial Z_i^{\gamma }}\frac 1{K_{\mu }}\sum _{j=1}^{K_\mu }\delta (Z_j^{\mu }-U).
\label{equa6}
\end{equation}

Clearly, $\frac{\partial }{\partial Z_i^{\gamma }}\sum _{j=1}^{K_{\mu }}\delta (Z_j^{\mu }-U)=0 \text{ if } \gamma \neq \mu $, so that $$\frac{\mathit{\partial }}{\mathit{\partial }Z_i^{\mu }}\sum _{j=1}^{K_{\mu }}\delta (Z_j^{\mu
}-U)=\frac{\partial }{\partial Z_i^{\mu }}\delta (Z_i^{\mu }-U),  i=1,2,\ldots, K_{\mu }.$$
Hence \eqref{equa6} can be rewritten
\begin{align}
\gamma^\mu _1&=\frac{-\mathit{\partial }}{\mathit{\partial }U}\left\{\int _{{R^{m+2}} }\prod _{\delta =1}^P\prod _{l=1}^{K_{\delta
}}dZ^{\delta }_l\sum _{i=1}^{K_{\mu }}\left\{ \mathcal{F}^{\mu }(U)p_t((Z_i^{\alpha })_{i,\alpha})\right\}\frac
1{K_{\mu }}\delta (Z_i^{\mu }-U)\right\} \\
&=\frac{-\partial }{\partial U}\left\{\int _{{R^{m+2}} }\prod _{\delta =1}^P\prod _{l=1}^{K_{\delta }}dZ^{\delta
}_l\left\{ \mathcal{F}^{\mu }(U)\widehat{n}{^{\mu }}   (U)p_t((Z_i^{\alpha })_{i,\alpha})\right\}\right\}.
\end{align}
Finally, using \eqref{expect},  $\gamma^\mu _1$  is given by
\begin {equation}
\gamma^\mu _1=\frac{-\mathit{\partial }}{\mathit{\partial }U}( \mathcal{F}^{\mu }(U)n^{\mu }(U,t)).
\label{equa7}
\end{equation}

\medskip
Let us now consider the second term  $\gamma^\mu _2$ in  \eqref{equa5}
\begin {equation}
\gamma^\mu _2=-\int _{\mathit{{R^{m+2}} }}\prod _{\delta =1}^P\prod _{l=1}^{K_{\delta }}dZ^{\delta }_l 
 \\
 \sum_{\gamma =1}^P\sum _{i=1}^{K_{\gamma }}\frac{\partial }{\partial Z_i^{\gamma }}\left\{\sum _{\alpha =1}^P\frac 1{K_{\alpha
}}\sum _{l=1}^{K_{\alpha }}\mathcal{M}^{\gamma \alpha }(Z^{\gamma }_i,Z^{\alpha }_l)p_t((Z_i^{\alpha })_{i,\alpha})\right\}\frac 1{K_{\mu }}\sum _{j=1}^{K_{\mu }}\delta (Z_j^{\mu }-U).
\label{equa8}
\end{equation}
By the same argument used for the derivation of \eqref{equa7}, an integration by parts in  \eqref{equa8}
implies
\begin{align}
\gamma^\mu _2&=\int _{\mathit{{R^{m+2}} }}\prod _{\delta =1}^P\prod _{l=1}^{K_{\delta }}dZ^{\delta }_l \,
 \sum_{\gamma =1}^P\sum _{i=1}^{K_{\gamma }}\left\{\sum _{\alpha =1}^P\frac 1{K_{\alpha }}\sum _{l=1}^{K_{\alpha
}}\mathcal{M}^{\gamma \alpha }(Z^{\gamma }_i,Z^{\alpha }_l)p_t((Z_i^{\alpha })_{i,\alpha})\right\}\frac{\partial }{\partial Z_i^{\gamma }}\frac 1{K_{\mu }}\sum _{j=1}^{K_{\mu }}\delta (Z_j^{\mu }-U) \\
&=-\frac{\partial }{\partial U}\int _{{R^{m+2}} }\prod _{\delta =1}^P\prod _{l=1}^{K_{\delta }}dZ^{\delta }_l
\sum _{i=1}^{K_{\mu
}}\sum _{\alpha =1}^P\frac 1{K_{\alpha }}\sum _{l=1}^{K_{\alpha }}\mathcal{M}^{\mu \alpha }(U,Z^{\alpha }_l)p_t((Z_i^{\alpha })_{i,\alpha})\frac 1{K_{\mu }}\delta (Z_i^{\mu }-U) \nonumber \\
&=-\frac{\partial }{\partial U}\int_{R^{m+2}} \mathit{dU}\mathit{^\prime}\int _{\mathit{{R^{m+2}} }}\prod _{\delta =1}^P
\prod_{l=1}^{K_{\delta }}dZ^{\delta }_l\sum _{i=1}^{K_{\mu }}\sum _{\alpha =1}^P\frac 1{K_{\alpha }}\frac 1{K_{\mu }}
\sum_{l=1}^{K_{\alpha }}\mathcal{M}^{\mu \alpha }(U,U^\prime) \,
p_t((Z_i^{\alpha })_{i,\alpha})\delta({Z}{_i^{\mu }}-U)\delta (Z_l^{\alpha }-U^\prime) \nonumber \\
&=-\frac{\partial }{\partial U}\int_{R^{m+2}} \mathit{dU}\mathit{^\prime}\int _{\mathit{{R^{m+2}} }}\prod _{\delta =1}^P\prod
_{l=1}^{K_{\delta }}dZ^{\delta }_l\sum _{\alpha =1}^P\mathcal{M}^{\mu \alpha }(U,U^\prime)p_t((Z_i^{\alpha })_{i,\alpha})\widehat{n}{^{\mu }}   (U)\widehat{n}{^{\alpha }}   (U^\prime)  \nonumber
\end{align}
which again leads to
\begin{equation}
\gamma^\mu _2=-\frac{\partial }{\partial U}\int_{R^{m+2}} \mathit{dU}\mathit{^\prime}\sum _{\alpha =1}^P\mathcal{M}^{\mu \alpha
}(U,U^\prime)\langle\widehat{n}{^{\mu }}   (U)\widehat{n}{^{\alpha }}    (U^\prime)\rangle_{p_t}.
\label{gamma_2}
\end{equation}

The term $\gamma^\mu _2$  has been obtained without the use of approximations and hence is exact. However, it is not really useful in applications. A way to go further consists in making the so called mean field estimate (see e.g.\blue{\cite{Kur1}})
\begin {equation}
\langle\widehat{n}{^{\mu }}(U)\widehat{n}{^{\alpha }}(U^\prime)\rangle_{p_t}\approx\langle\widehat{n}{^{\mu }}(U)\rangle
 _{p_t}\langle\widehat{n}{^{\alpha }}(U^\prime)\rangle_{p_t}.
\end{equation}

This approximation is valid because the fluctuations of  $\widehat{n}{^{\mu }}(U)$ (resp.  $\widehat{n}{^{\alpha }}(U^\prime)$) are small for  $K_{\mu }$ (resp.  $K_{\alpha }$) large and are $O(\frac 1{\sqrt{K_{\mu }}})$ (resp.  $O(\frac 1{\sqrt{K_{\alpha }}})$).

Accordingly, the coupling term  $\gamma^\mu _2$ takes the form :
\begin {equation}
\gamma^\mu _2=-\frac{\mathit{\partial }}{\mathit{\partial }U}\int_{R^{m+2}} \mathit{dU}\mathit{^\prime}\sum _{\alpha=1}^P\mathcal{M}^{\mu \alpha }(U,U^\prime)n^{\mu }(U,t)n^{\alpha }(U^\prime,t).
\end{equation}
The last diffusive term  $\gamma^\mu _3$ can be derived in a similar fashion. It is given by
\begin {equation}
\gamma^\mu _3=\frac{1}{2}\int _{\mathit{{R^{m+2}} }}\prod _{\delta =1}^P\prod _{l=1}^{K_{\delta }}dZ^{\delta }_l
 \\
 \sum
_{\gamma =1}^P\sum _{i=1}^{K_{\gamma }}(\beta _V^{\gamma})^2 \frac{\partial ^2}{\partial (V_i^{\gamma})^ 2}\,p_t((Z_i^{\alpha })_{i,\alpha
})\frac 1{K_{\mu }}\sum _{j=1}^{K_{\mu }}\delta (Z_j^{\mu }-U).
\end{equation}
Integration by parts gives
\begin {equation}
\gamma^\mu _3=\frac{1}{2}\int _{\mathit{{R^{m+2}} }}\prod _{\delta =1}^P\prod _{l=1}^{K_{\delta }}dZ^{\delta }_l
 \\
 \sum
_{\gamma =1}^P\sum _{i=1}^{K_{\gamma }}(\beta _V^{\gamma})^2 p_t((Z_i^{\alpha })_{i,\alpha
})\frac 1{K_{\mu }}\frac{\partial ^2}{\partial (V_i^{\gamma})^ 2 }\sum _{j=1}^{K_{\mu }}\delta (Z_j^{\mu }-U).
\end{equation}
Here also, one has 
\begin {equation}
\frac{\partial ^2}{\partial (V_i^{\gamma})^ 2}\sum _{j=1}^{K_{\mu }}\delta (Z_j^{\mu }-U)=0 \text{ if } \gamma \neq \mu
\end{equation}
while 
\begin {equation}
\frac{\partial ^2}{\partial (V_i^{\mu})^ 2}\sum _{j=1}^{K_{\mu }}\delta (Z_j^{\mu }-U)=\frac{\partial ^2}{\partial (V_i^{\mu})^ 2}\delta (Z_i^{\mu }-U).
\end{equation}
Thus
\begin {equation}
\gamma^\mu _3=\frac{1}{2}\int _{\mathit{{R^{m+2}} }}\prod _{\delta =1}^P\prod _{l=1}^{K_{\delta }}dZ^{\delta }_l
 \\
p_t((Z_i^{\alpha })_{i,\alpha
})\frac 1{K_{\mu }}\sum_{i=1}^{K_\mu}(\beta _V^{\mu})^2 \frac{\partial ^2}{\partial (V_i^{\mu})^ 2}\delta (Z_i^{\mu }-U).
\end{equation}
So
\begin {equation}
\gamma^\mu _3=\frac{(\beta _V^{\mu})^2}{2}\frac{\partial ^2}{\partial u^2}\int _{\mathit{{R^{m+2}} }}\prod _{\delta =1}^P\prod _{l=1}^{K_{\delta }}dZ^{\delta }_l
 \\
p_t((Z_i^{\alpha })_{i,\alpha
})\frac 1{K_{\mu }}\sum_{i=1}^{K_\mu}\delta (Z_i^{\mu }-U)
\end{equation}
and hence
\begin {equation}
\gamma^\mu _3=\frac{(\beta _V^{\mu})^2}{2}\frac{\partial ^2}{\partial u^2}n^{\mu}(U,t).
\end{equation}

Finally, the neural population probability distributions $n^{\mu}(U,t)$ which are the expectation values of the set variables $\widehat{n}{^{\mu }}(U)\; \mu=1,2, \ldots, P$ satisfy the following system of (nonlinear) $\mathrm{IPDE}$
\begin {equation}
\frac{\mathit{\partial }}{\mathit{\partial }t}n^{\mu }(U,t)=-\frac{\partial }{\partial U}( \mathcal{F}^{\mu }(U)n^{\mu
}(U,t))-\frac{\partial }{\partial U}\int_{R^{m+2}} \mathit{dU}\mathit{^\prime}\sum _{\alpha =1}^P\mathcal{M}^{\mu \alpha }(U,U^\prime)n^{\mu
}(U,t)n^{\alpha }(U^\prime,t)+
\frac 1 2(\beta _V^{\mu})^2 \frac{\partial ^2}{\partial U^2}n^{\mu }(U,t)
\label{generalPDE}
\end{equation}

\begin {equation*}
\mu =1,2,\dots, P, \ \;U=(u,\widehat{z},s), \ u,s\in \mathbb{R}, \;\widehat{z} \in \mathbb{R}^m, \; U^\prime=(u^\prime,\widehat{z^\prime},s^\prime), \ u^\prime,s^\prime \in \mathbb{R}, \;\widehat{z^\prime} \in \mathbb{R}^m.
\end{equation*}
Let us recall $U=(U_l)_{l=1,2, \ldots, m+2}$. We write explicitly the components of $U$ as $U=(u,\widehat{z},s), \; u,s\in \mathbb{R}, \;\widehat{z} \in \mathbb{R}^m$, where u (resp. $\widehat{z},s$) has potential (resp. recovery, synaptic) meaning.

We have thus obtained a system of equations for the study of interacting large-scale neural populations using a mean field approach. 
The dynamical behavior of the neurons in these populations is developed in terms of HH concepts which are usually capable of giving a quantitative description of ionic currents across the neural membranes. Interneuronal connections in these networks are not necessarily constant over time. A synaptic plasticity mechanism is proposed (see \blue{\cref{Isyn,Jij,Omega}}). For such systems, equation\eqref{generalPDE}, which describes the evolution over time of the PPD for the population $\mathscr{P}_{\mu}$  which interacts with populations $\mathscr{P}_{\nu}, \nu = 1,2,\dots, P$ is of the same type as that obtained in\blue{\cite{ToubFaug}} where it was shown, using methods of probability theory, that in the limit of large populations, the network may display the property of  propagation of chaos.  
In such a way, all neurons which belong to a given population, asymptotically have the same probability distribution which characterizes a mean field stochastic process. The equation \eqref{generalPDE}  is the (MVFP) equation for this  process.

\section{Application to  coupled large-scale Fitzhugh-Nagumo networks}
\label{sec:applFN}

\subsection{Population probability distributions for coupled Fitzhugh-Nagumo networks}
\label{subsec:popdensFN}
In what follows, we give an illustration of the general result \eqref{generalPDE} in the case of FN model neurons which  leads to a simple example of the stochastic system \eqref{syststochgenVect}. The basic excitability properties of many neurons are
approximately  described by the FN model and the mathematical tractability of the resultant system  makes it suitable as a first application of the theory. 
 
The dynamical  system is now given by (for all $i=1,2,\ldots, K_{\gamma }, \gamma =1,2,\ldots, P$)
\begin{align}
\frac{dV_i^{\gamma}}{dt} =\ & \mathit{F}^{\gamma }(V_i^{\gamma },X_i^{\gamma})+I_{\mathit{ext}}^{\gamma }(t)+\eta_{i,t}^{\gamma }+\sum _{\alpha =1}^P\frac 1{K_{\alpha }}\sum_{j=1}^{K_{\alpha }}M^{\gamma \alpha }(V_i^{\gamma },X_i^{\gamma },\phi_i^{\gamma },V_j^{\alpha },X_j^{\alpha },\phi_j^{\alpha })
\label{FiNa0} \\
\frac{dX_i^{\gamma}}{dt} =\ &\mathit{G}^{\gamma }(V_i^{\gamma },X_i^{\gamma }) \label{FiNa0a} \\ 
\frac{d \phi_i^{\gamma}}{dt} =\ & \Omega^{\gamma}(\phi_i^{\gamma},V_i^{\gamma},X_i^{\gamma}) \label{FiNa0b}
\end{align}
where
\begin {align}
\mathit{F}^{\gamma }(V,X)=\ &-k^{\gamma }V(V-a^{\gamma })(V-1)-X
\label{form1} \\
\mathit{G}^{\gamma }(V,X)=\ &b^{\gamma }(V-m^{\gamma }X).
\label{form2}
\end{align}
The variables $V_i^\gamma$ represent the membrane potential of the $i^{th}$ cell in $\mathscr{P}_{\gamma}$ 
 while, in FN networks, the variables  $\hat R_i^\gamma$ which have been introduced in \eqref{vars} are  called $X_i^\gamma$ and are one-dimensional.
The parameters $k^{\gamma },a^{\gamma },b^{\gamma},m^{\gamma }$  govern the dynamics of the FN  neural model in the population $\mathscr{P}_{\gamma}$  (see e.g.\blue{\cite{Tuck1}} for typical values of these parameters). 
The quantities $I_{\mathit{ext}}^{\gamma }(t)$, $\mathit{\eta}_{i,t}^{\gamma }$ have the same meaning as in section~\bref{subsec:DynSystOnePop}.

In the following sections, we shall give a more precise form for the coupling term $M^{\gamma \alpha}(V_i^{\gamma },X_i^{\gamma },\phi_i^{\gamma },V_j^{\alpha },X_j^{\alpha },\phi_j^{\alpha })$  in the case of uniformly and non uniformly distributed sets of connection parameters between the cells of these populations.
According to the general result \eqref{generalPDE}, in the mean field limit, the $\mathrm{PPD}$s $n^{\mu }(V,X,\phi,t)$ \ $ \mu =1,2,\ldots, P$, which are built for a set of interacting populations of FN neurons, satisfy the following system of (nonlinear) integro-partial differential equations 
\begin{align}
\label{equa9}
&\frac{\mathit{\partial }}{\mathit{\partial }t}n^{\mu }(V,X,\phi,t)=\\
&-\frac{\partial }{\partial V}(( F^\mu(V,X)+I_{ext}^\mu(t))n^{\mu}(V,X,\phi,t))-\frac{\partial }{\partial X}( G^\mu(V,X)n^{\mu}(V,X,\phi,t))  -\frac{\partial }{\partial \phi}( \Omega^\mu(\phi,V,X)n^{\mu}(V,X,\phi,t)) \nonumber  \\
&-\frac{\partial }{\partial V} \int_{R^{3}} \mathit{dV}\mathit{^\prime}\mathit{dX}\mathit{^\prime}\mathit{d\phi}\mathit{^\prime}\sum _{\alpha =1}^P{M}^{\mu \alpha }(V,X,\phi,V^\prime,X^\prime,\phi^\prime)n^{\mu}(V,X,\phi,t)n^{\alpha }(V^\prime,X^\prime,\phi^\prime,t) \nonumber \\
&+\frac 1 2(\beta _V^{\mu})^2 \frac{\partial ^2}{\partial V^2}n^{\mu }(V,X,\phi,t) \qquad \mu =1,2,\dots ,P. \nonumber
\end{align}

We shall give examples of applications of this mean-field  approach to neural networks dynamics by considering the cases of firstly one population and then two
populations of FN cells. Equation \eqref{equa9} will be solved  numerically and its solutions  compared to the numerical solutions of the stochastic
system \blue{\cref{FiNa0,FiNa0a,FiNa0b}}.

\subsection{Statistical measures from the stochastic and integro-partial differential systems}
\label{subsec:measPIDE}

Our goal is to have, at least numerically, a reasonably good understanding of the dynamics of the above considered set of  neural populations.
From the solution of the system of 
$\mathrm{IPDE}$  \eqref{equa9}, it is  in principle possible to have insights into  this dynamics, in the limiting case of
infinitely many cells. 
Since the real situation corresponds generally to a large  (but finite) number of cells, the 
solutions of  \eqref{equa9}  \ may be viewed only as an approximation.

Another approach for the study of the network dynamical behavior is based on
a direct numerical evaluation of solutions of the (stochastic) coupled differential equations \blue{\cref{FiNa0,FiNa0a,FiNa0b}}. 
Let us call this system $\mathrm{SCDE}$. 
In \blue{\cite{Brette}}, one can find a review of various methods of numerical solution of these systems, and in\blue{\cite{Djurfeldt}} are given appropriate techniques for solving such very large systems.  

Clearly, the solution of $\mathrm{IPDE}$  has a great advantage over the direct solutions of the  $K_{\gamma }$  dimensional
 $\mathrm{SCDE}$, $\gamma =1,2,\ldots, P$ since only $P$ equations must be solved.

Both systems ($\mathrm{IPDE}$ and $\mathrm{SCDE}$) have been numerically implemented over the same domain $\Omega\times\Lambda\times\Phi$, where $\Omega =[V_{min}, V_{max}]$ (resp.  $\Lambda =[X_{min}, X_{max}]$, $\Phi =[\phi_{min}, \phi_{max}]$) are the bounded domains of variation of potential (resp.  recovery, synaptic coupling variables).

Moreover, similar initial conditions have been chosen in both cases. More precisely, a normalized Gaussian function $n(V,X,\phi,t=0)$ has been considered for $\mathrm{IPDE}$ with mean and standard deviation parameters $(V_0,\sigma_V), (X_0,\sigma_X), (\phi_0,\sigma_{\phi})$.
For $\mathrm{SCDE}$,  all cells initial data $(V_i^{\gamma}(t=0), X_i^{\gamma}(t=0),\phi_i^{\gamma}(t=0)), i=1,2,\ldots, K_{\gamma }, \gamma =1,2,\ldots, P$ were selected through the use of $i.i.d.$ Gaussian random variables with the same moments as the ones used for the solution of $\mathrm{IPDE}$.

Concerning the numerical aspects of the problem, a discretization procedure is necessary for $\mathrm{IPDE}$, requiring a partition of the parameter space $\Omega \times \Lambda \times  \Phi$. 
Obviously, this method of solution takes much less computing time than the possibly large numbers of trials required for SCDE.  

In order to compare the results obtained by the solutions of $\mathrm{IPDE}$ and $\mathrm{SCDE}$,  it is necessary to show how one can extract information  
on each dynamical variable $V, X$ or $\phi$, for any cell in the full system, at any time, from a probabilistic point of view.
If one  concentrates on the potential variable, for example, a neural population potential distribution has to be defined.

Concerning $\mathrm{IPDE}$, we have at our disposal the $\mathrm{PPD}$ $n^{\mu }(V,X,\phi,t)$, solution of  \eqref{equa9} and this potential distribution can be defined  as the marginal 
\begin {equation}
\rho^{\mu }(V,t)=\int_{R^2}n^{\mu }(V,X,\phi,t) \, dX d\phi.
\label{ro}
\end{equation}
In that way, any statistical measure may be obtained through  $\rho ^{\mu }(V,t)$. 
In the following, we call $\boldlangle  V(t) \boldrangle _{\distrho}$, $\mathit{Var}(V(t))_{\distrho}$ the mean and variance of the membrane potential with respect to $\rho^\mu$.

We define a firing measure  $\varphi_{\!\distrho}(t)$, as the probability that, at a given time $t$, the potential  $V(t)>\theta $ where  $\theta $ is some prescribed threshold
\begin {equation}
\varphi_{\!\distrho}(t)=\int _{\theta }^{+\infty}\rho ^{\mu }(\apos{V},t)\mathit{d\apos{V}}.
\label{Fir1}
\end{equation}

Regarding $\mathrm{SCDE}$, we propose a numerical measure of the $\mathrm{PPD}$ in the following way. Rather than following the trajectories of individual neurons (each starting from initial conditions $(V,X,\phi)_i^{\mu}(t=0), i=1,2,\ldots, K_{\mu}, \mu =1,2,\ldots, P$), we set up a counting process about the presence (at a given time  $t$, for any cell  $i$ in  $\mathscr{P}_{\mu }$) of  the potential $V_i^{\mu}(t)$   in a given subinterval $[a_j, b_j]=\text{bin}_j$. The intervals  $[a_j,b_j], j=1,2,\ldots, N$ build a partition of  $\Omega $, all having the same length $\lambda =(V_{\mathit{max}}-V_{\mathit{min}})/N$ (see section \bref{sec:Num} for domain extensions). 

Because of the presence of noise, one has to perform sufficiently many trials. Accordingly, the $\mathrm{PPD}$ for $\mathrm{SCDE}$ will be defined as the discrete distribution  $\xi^{\mu}(t)$ given by
\begin {equation}
\xi^{\mu}(t) = \{ \, p_j^{\mu }(t), \ \ j=1,2,\ldots, N \, \} = \bigg\{ \, \bigg( \frac{\mathcal{N}(V_i^{\mu }(t)\in [a_j,b_j])}{K_{\mu }\ast
\mathit{N_{trials}}},\ 1<i<K_{\mu } \bigg), \ j=1,2,\ldots, N \, \bigg\}
\label{rhodisc}
\end{equation}
$\mathcal{N}$ denotes the number of occurrences for all trials and $\mathit{N_{trials}}$ the total number of trials which have been used in the simulations. Clearly, $\xi^{\mu}(t)$ is a piecewise constant function, each $p_j^{\mu }(t)$ giving the average number of events in each  $\text{bin}_j,\, j=1,2,\ldots, N$.\\
Therefore, for $\mathrm{SCDE}$, the mean and variance of $V(t)$ with respect to this discrete probability are given by
\begin{align}
\boldlangle V(t) \boldrangle_{\distxi}\; =\ & \sum _{j=1}^N(V_{\mathit{min}} + \lambda j)\,p_j^{\mu }(t) 	\label{Mo2} \\
\mathit{Var}(V(t))_{\distxi} \; =\ & \sum _{j=1}^N\, (V_{\mathit{min}} + \lambda j)^2\, p_j^{\mu }(t)-
\big(\sum_{j=1}^N\, (V_{\mathit{min}} + \lambda j)\, p_j^{\mu }(t)\big)^2.						\label{Va2}
\end{align}
Finally, we evaluate a discrete firing measure  $\varphi_{\distxi}(t)$ as the probability that the potential $V(t)>\theta$ 
\begin {equation}
\varphi_{\distxi}(t) = \sum _{j\,>\,(\theta -V_{\mathit{min}})/{\lambda}} p_j^{\mu }(t).
\label{Fir2}
\end{equation}

In the next section, we compare the two types of probability distributions, the continuous one  $\rho ^{\mu }(V,t)$ and
the discrete one $\xi^{\mu}(t)$, as well as averages, standard deviations and firing measures  derived from these two distributions.

\subsection{Numerical results for a single population}
\label{subsec:numres1pop}
Let us first give the form of the equation \eqref{equa9} in the case of only one population, namely $P=1$.
In the system \blue{\cref{FiNa0,FiNa0a,FiNa0b}},
the matrix $M^{\gamma \alpha }(V_i^{\gamma },X_i^{\gamma},\phi_i^{\gamma},V_j^{\alpha },X_j^{\alpha },\phi_j^{\alpha })$ reduces to the term $M^{11}(V_i^1,X_i^{1},\phi_i^{1},V_j^1,X_j^1,\phi_j^1)), \, i=1, 2,\ldots, K_1$. The triples $(V_i^1,X_i^1,\phi_i^1)$ are rewritten $(V_i,X_i,\phi_i)$ and $M^{11}$, $K_1$, $n^1(V,X,\phi,t)$ are rewritten $M$, $K$ and $n(V,X,\phi,t)$, respectively.   
Similarly, the densities $\rho^{1}(V,t)$ and $\xi^{1}(t)$ will be denoted by  $\rho(V,t)$ and $\xi(t)$.

\subsubsection {Excitatory synapses, uniform connectivity}
\label{subsubsec:ExcInhSynUnif}

In the present section, we assume that the population has all to all connectivity as in\blue{\cite{Buice2,Fasoli}}.
For discussions on this topic see\blue{\cite{Amit,Fasoli2,Tuck5}}.

We give here an explicit form of the synaptic current $I_{syn,i}(t)$ introduced in \eqref{Isyn}, in the case where all cells in the network are connected with excitatory synapses.
Because of the homogeneous character of the coupling considered here, the variables $\phi_{i}, i=1,2,\ldots, K$ assume constant values. We call $J^{\mathit{E}}$  the constant coupling term $\Gamma(\phi_i, \phi_j)$ (see \eqref{Jij}) and $V_S^{\mathit{E}}$ the corresponding parameters for resting values.
Moreover, a simple form has been considered for the detection of presynaptic events using sigmoidal functions $\sigma^{\mathit{E}}_{\Theta}$, defined by
\begin{equation}
\sigma^{\mathit{E}}_{\Theta}(V)=(1+e^{-\beta(V-V^{\mathit{E}}_{\Theta})})^{-1}
\label{sigmoidV}
\end{equation}
for some prescribed threshold values $ V^{\mathit{E}}_{\Theta}$, and parameter $\beta$. One has
\begin{equation}
M(V_i,X_i,V_j,X_j)=J^{\mathit{E}}(V_S^{\mathit{E}}-V_i)\sigma^{\mathit{E}}_{\Theta}(V_j)
\label{couplop}
\end{equation}
where the indices $i$ (resp. $j$) refer to post (resp. pre) synaptic cells. Finally, the synaptic input current $I_{syn,i}(t)$ for each cell $i$ is given by 
\begin{equation}
I_{syn,i}(t)=\frac{1}{K} \sum_{j=1}^K M(V_i,X_i,V_j,X_j).
\label{synapticcells}
\end{equation}
Accordingly, the stochastic system ($\mathrm{SCDE}$) \blue{\cref{FiNa0,FiNa0a,FiNa0b}} takes the form
\begin {equation}
\begin{aligned}
\frac{d V_i}{dt}=\ &\mathit{F}(V_i,X_i)+I_{\mathit{ext}}(t)+\eta_{i,t}+I_{syn,i}(t) \\
\frac{dX_i}{dt} =\ &\mathit{G}(V_i,X_i) \qquad\qquad i=1, 2,\ldots, K
\end{aligned}
\label{FiNa3}
\end{equation}
while the integro-partial differential equation ($\mathrm{IPDE}$) \eqref{equa9} for the $\mathrm{PPD}$ $n(V,X,t)$ is
\begin{align}
\frac{\mathit{\partial }}{\mathit{\partial }t}\, n(V,X,t)&=\frac{-\partial }{\partial V}\left\{(\mathit{F}(V,X)+I_{\mathit{ext}}(t))\, n(V,X,t)\right\}-\frac{\partial }{\partial X}\left\{\mathit{G}(V,X)\, n(V,X,t)\right\}
\label{PDE2} \\
&-J^{\mathit{E}}\frac{\mathit{\partial }}{\mathit{\partial }V}\left\{(V_S^{\mathit{E}}-V)\, n(V,X,t)\right\}\int
_{R^2}\mathit{dV'}\mathit{}\mathit{dX}\mathit{^\prime}\sigma^{\mathit{E}}_{\Theta}(V^\prime)\, n(V^\prime,X^\prime,t) \nonumber 
+\frac{\beta _V^2} 2\frac{\partial ^2}{\partial V^2}\, n(V,X,t). \nonumber
\end{align}
Numerical integration of \eqref{FiNa3}, \eqref{PDE2} have been performed, with suitable initial conditions. The results are shown in Fig.\,\blue{\ref{ExciInh}} below. 
A 3--dimensional graph of the time behavior of the discrete probability density $\xi(t)$ is represented in Fig.\,\blue{\subref*{discprob}} 
which may be compared to the behavior of the continuous one $\rho(V,t)$ shown in Fig.\,\blue{\subref*{contprob}}. 

\begin{figure}[[!h]
\centering
\subfloat[Mean values of potential obtained from $\mathrm{SCDE}$ and $\mathrm{IPDE}$.
]
{\includegraphics[width=0.45\textwidth]{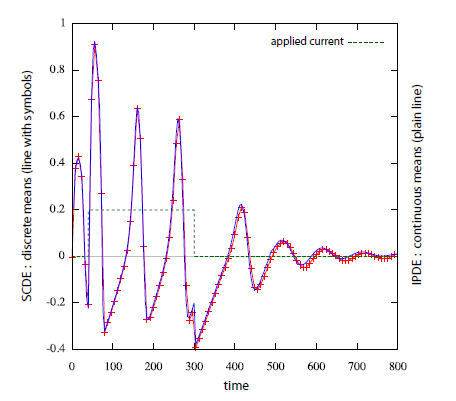} \label{Exci-a}}
\hfill
\subfloat[Variance  of potential obtained from $\mathrm{SCDE}$ and $\mathrm{IPDE}$.
]
{\includegraphics[width=0.45\textwidth]{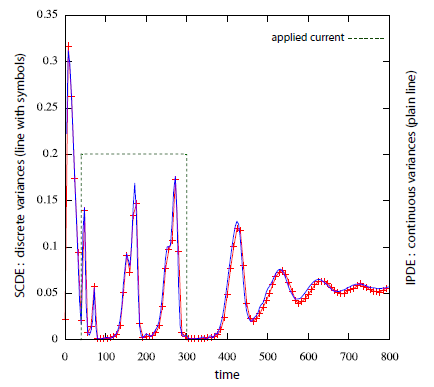} \label{ExciInh-d}}
\hfill
\subfloat[Depolarizations versus time for 5 randomly selected cells ($\mathrm{SCDE}$).]
{\includegraphics[width=0.5\textwidth]{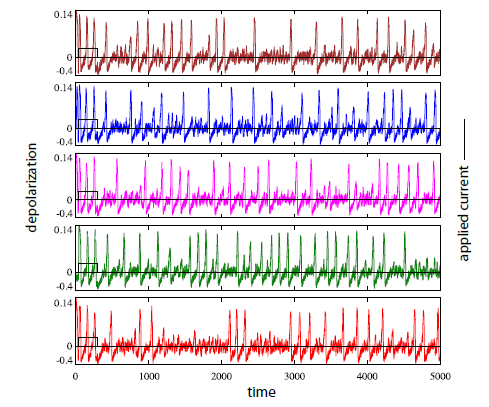} \label{Exci-b}}
\captionsetup{justification=centering}
\caption{Excitatory synapses, uniform connectivity.\\ Parameters for the FN system and noise, see section  \bref{sec:Num}. }
\label{Exci}
\end{figure}

Similarly, in Fig.\,\blue{\subref*{Exci-a}} and Fig.\,\blue{\subref*{ExciInh-d}}, the mean values and variances of the potential variables obtained from the solutions of $\mathrm{SCDE}$ and $\mathrm{IPDE}$, are compared. These quantities give a global description of the network activity. It is seen that there is close agreement between the two methods of solution.

\begin{figure}[!h]
\centering
\subfloat[Evolution of the normalized discrete probability density $\xi(t)$, solution of the stochastic dynamical system  \eqref{couplop}--\eqref{FiNa3}.]
{\includegraphics[height=0.33\textwidth]{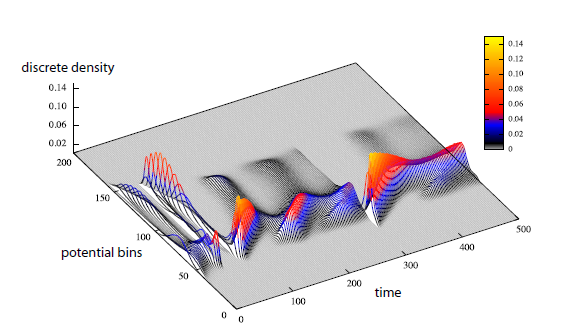} \label{discprob}} \\
\subfloat[Evolution of the normalized continuous probability density $\rho(V,t)$, solution of the MVFP equation \eqref{PDE2}.]
{\includegraphics[height=0.33\textwidth]{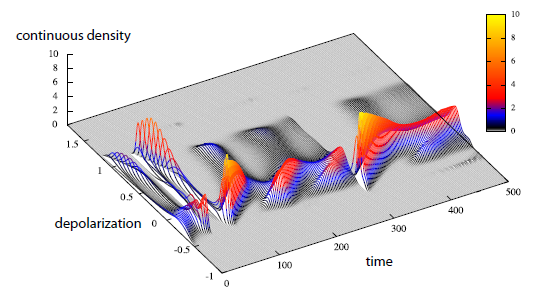} \label{contprob}} \\
\caption{Time behavior of  probability densities. Initial conditions and simulation parameters, see section \bref{sec:Num}.}
\label{ExciInh}
\end{figure}

In addition, one can obtain insight into the activity of cells from a local point of view. This  is illustrated in Fig.\,\blue{\subref*{Exci-b}} showing the superimposed time development, over several trials, of the potential of some arbitrarily chosen cells in the network.
There appears significant variability in the activity of individual cells in the network. However, as shown in Fig.\,\blue{\subref*{Exci-a}}, an order emerges in the mean behavior due to synchronizations between cells.

\subsubsection{Excitatory synapses, time dependent non uniform connectivity}
\label{subsubsec:ExcSynNonUnif}
In section  \bref{subsubsec:ExcInhSynUnif}, we considered the case of one population with uniform and constant connections over time. In this section, we present the structure of the mean field equation and the nature of the  numerically obtained solutions, in the case of one population (i.e. $P=1$)  whose non-uniform interconnections can vary over time.

\begin {comment}
The functions introduced in \eqref{Jij} and \eqref{Omega} in the definition of the coupling  have the following form
\begin{align}
\mathit{ \Gamma}(\phi_i, \phi_j) &=\phi_i \, \phi_j    \\ 
\mathit{\Omega}(\phi_i,X_i) &=-\alpha \, \phi_i+ \beta \, \sigma_{\phi}(X_i) 
\label{gammaform}
\end{align}
where $\alpha$ and $\beta$ are some parameters and  $\sigma_{\phi}$ is a sigmoidal function, $\sigma_{\phi}(X)=(1+e^{-\nu(X-X_{S})})^{-1}$, for some prescribed threshold value $X_{S}$ and parameter $\nu$. $\hfill\refstepcounter{equation}(\theequation) \label{sigmoidX}$
 \end{comment}

In section \bref{subsec:DynSystOnePop}, the functions $\Gamma$ and $\Omega$ have been introduced for the definition of the coupling terms. 
As in section \bref{subsubsec:ExcInhSynUnif}, the dynamical behavior of synaptic excitatory conductances is viewed 
in terms of the sigmoidal dependence $\sigma^{\mathit{E}}_{\Theta}(V)$ with respect to presynaptic depolarization. 
The connection term $M$ (see \eqref{synapticcells}), which we consider now, is
$M(V_i,X_i,\phi_i, V_j,X_j,\phi_j) =  (V^{\mathit{E}}_S-V_i)\, {J^{\mathit{E}}} \, \Gamma (\phi_i \,, \phi_j \,) \, \sigma^{\mathit{E}}_{\Theta}(V_j)
$ where $V^{\mathit{E}}_S$ and $J^{\mathit{E}}$ are constant.
Thus, the postsynaptic current which is applied to each cell $i=1,2,\dots, K$ is
\begin{equation}
I_{syn,i}(t) = (V^{\mathit{E}}_S-V_i) \frac{J^{\mathit{E}}}{K} \sum_{j=1}^K\Gamma (\phi_i \,, \phi_j \,) \sigma^{\mathit{E}}_{\Theta}(V_j) .
\label{IsynExcNonUnif}
\end{equation}
Finally, the structure of the network dynamical system in the case of non uniform, time dependent couplings between cells, which we wish 
to analyze from the kinetic point of view, is the following
\begin{equation}
\begin{aligned}
\frac{d V_i}{dt} =\ &\mathit{F}(V_i,X_i)+I_{\mathit{ext}}(t)+\eta_{i,t} +I_{syn,i}(t)\\ 
\frac{d X_i}{dt} =\ &\mathit{G}(V_i,X_i) \\
\frac{d\phi_i}{dt} =\ &\Omega(\phi_i,X_i).
\label{inhomecha}
\end{aligned}
\end{equation}

The general result \eqref{generalPDE} can now be applied to produce the structure of the mean field equation which can be derived for this kind of neural network system. One looks for the probability distribution $n(V,X,\phi, t)$ of depolarization, recovery and connection variables $(V,X,\phi) \in \mathbb{R}^3$ at time $t$. 
From \eqref{generalPDE},  $n(V,X,\phi, t)$ must satisfy
\begin{align}
\frac{\mathit{\partial }}{\mathit{\partial }t}n(V,X,\phi,t) & = \frac{-\partial }{\partial V}\left\{(\mathit{F}(V,X)+I_{\mathit{ext}}(t))\, n(V,X,\phi ,t)\right\}-\frac{\partial }{\partial X}\left\{\mathit{G}(V,X)\, n(V,X,\phi,t)\right\}      				\\           
& - J^{\mathit{E}}\frac{\partial }{\partial V}\left\{(V^{E}_S-V)\, \, n(V,X,\phi,t)\right\}\int_{R^3}\mathit{dV}\mathit{^\prime}\mathit{dX}\mathit{^\prime}\mathit{d\phi}\mathit{^\prime}\sigma^{E}_{\Theta} (V^\prime)\, \Gamma(\phi,\phi^\prime) \, n(V^\prime,X^\prime,\phi^\prime,t) \nonumber 	\\
& - \frac{\partial }{\partial \phi}\left\{\mathit{\Omega}(\phi,X)\, n(V,X,\phi,t)\right\} 
+\frac{\beta _V^2} 2\frac{\partial ^2}{\partial V^2}n(V,X,\phi,t).    \nonumber
\end{align}
Our approach here is the same as in previous sections where results obtained for $\mathrm{SCDE}$ and $\mathrm{IPDE}$  have been compared. The new variables $\mathit{\phi_i}, i=1,2,\dots, K $ have random character, in the same way as potential $V_i$ and recovery variables $X_i$. As has been described for potential variables in section \bref{subsec:measPIDE}, statistical measures for variables $\mathit{\phi_i} $ can be defined in both cases. 
For $\mathrm{IPDE}$, the probability distribution of $\phi$ variables is given by the marginal   
\begin {equation}
\chi(\phi,t)=\int _{\mathbb{R}^2}n(V,X,\phi,t)\mathit{dV}\mathit{dX}.
\label{chi}
\end{equation}
Thus, for example,  the (continuous) mean of the variable $\phi$ is obtained from 

\begin{equation}
\boldlangle \phi(t) \boldrangle_{\distchi}  =  \int _{\mathbb{R} }\mathit{\phi}\mathit{^\prime} \,  \chi(\mathit{\phi}\mathit{^\prime},t)\, \mathit{d\phi}\mathit{^\prime}.
\label{MoPhi}
\end{equation}

Concerning  $\mathrm{SCDE}$, the discrete version for the probability distribution of $\phi$-variables and related moments is obtained following the same lines as those developed for membrane potential  variables (see formulas \blue{\cref{rhodisc,Mo2,Va2,Fir2}}).  For all simulations which are described in this section, the function $\Omega$ has been taken to be of the form 
 \begin{equation}
 \mathit{\Omega}(\phi,X) =-\alpha_{\phi} \, \phi+ \beta_{\phi} \, \sigma_{\phi}(X)
 \label{OmegaFunc}
 \end{equation}
 where $\alpha_{\phi}$ and $\beta_{\phi}$ are 
 parameters and   $\sigma_{\phi}$ is a sigmoidal function
  \begin{equation}
\sigma_{\phi}(X)=(1+e^{-\nu(X-X_{thresh})})^{-1}
 \label{sigma}
 \end{equation}
  for some prescribed threshold value $X_{thresh}$ and parameter $\nu$. 
The function $\Gamma$ appearing in \eqref{IsynExcNonUnif} has been taken  to be of the Hebbian form 
\begin{equation}
\Gamma (\phi_i,\phi_j)=\phi_i \phi_j,
\label{Gamma1}
\end{equation}

\begin{figure}[!ht]
\centering
\subfloat[Excitatory synapses. Mean values of potential and connection variables obtained from $\mathrm{IPDE}$ and $\mathrm{SCDE}$.
No applied current.]
{\includegraphics[width=0.45\textwidth]{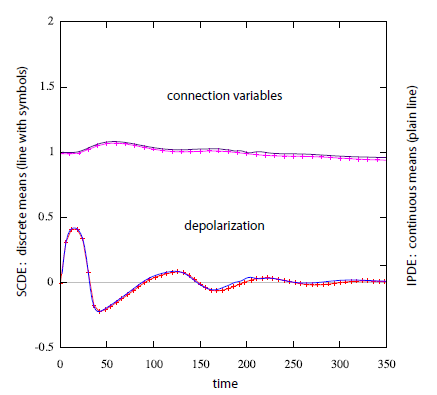}		\label{NonUnif-a}} 	\hfill
\subfloat[Excitatory synapses. Mean values of potential and connection variables obtained from $\mathrm{IPDE}$ and $\mathrm{SCDE}$. 
A step current is applied: rhythm generation.]
{\includegraphics[width=0.45\textwidth]{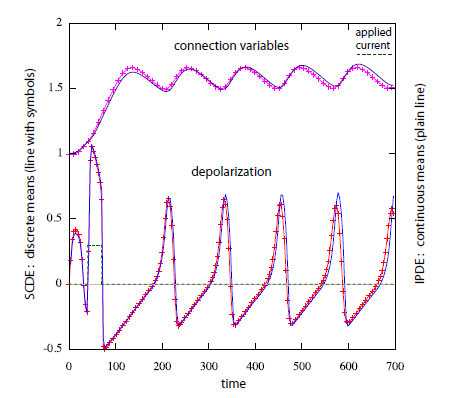}	\label{NonUnif-b}}   	\\
\subfloat[Excitatory synapses. Superimposed time development, over a sample of 5 trials, of potential and strength variables of 2 arbitrarily chosen cells.] 
{\includegraphics[width=0.45\textwidth]{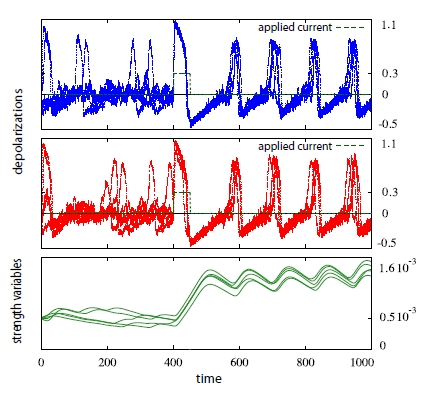}	\label{NonUnif-c}}           \hfill
\subfloat[Inhibitory synapses. Superimposed time development, over a sample of 5 trials, of potential and strength variables of 2 arbitrarily chosen cells .]
{\includegraphics[width=0.45\textwidth]{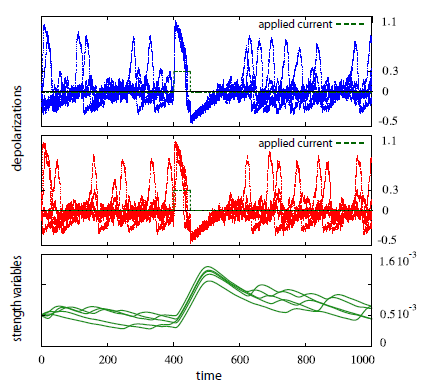}	\label{NonUnif-d}}   
\caption{Non uniform, time dependent connectivity. In Figs.3(c) and 3(d), the strength variables are defined as $J_{i_1 i_2}=\Gamma (\phi_{i_1},\phi_{i_2})$ (see  \eqref{Gamma1}) for 2 arbitrarily chosen cells, $\mathit{Cell}_{i_1}$ and $\mathit{Cell}_{i_2}$, $i_1\ne i_2 \in \{1...K\}$. 
For parameters of  FN system and noise, see section \bref{sec:Num}.}
\label{NonUnif} 
\end{figure}

   In Figs.\,\blue{\subref*{NonUnif-a}} and \blue{\subref*{NonUnif-b}}  are shown the mean values $\boldlangle V(t)\boldrangle_{\xi}$, $\boldlangle \phi(t) \boldrangle_{\xi}$ according to $\mathrm{SCDE}$ and $\boldlangle V(t) \boldrangle_{\rho}$, $\boldlangle \phi(t) \boldrangle_{\distchi}$ according to $\mathrm{IPDE}$, in the absence and in the presence of external current, for a network with excitatory connections. 
When there is no applied input current,  $\boldlangle V(t)\boldrangle_{\xi}$ and $\boldlangle V(t) \boldrangle_{\rho}$ decrease after a sufficiently long time. 
If the coupling parameter takes increasing values, the network activity is increasingly synchronized, even in the absence of a stimulus. 
The proposed mechanism  \eqref{inhomecha}, together with \eqref{IsynExcNonUnif} and \eqref{Gamma1} leads to significantly increased values
of these couplings, especially when an external current is applied to the network during a short period.
This is illustrated in Fig.\,\blue{\subref*{NonUnif-b}} where we see the emergence of a rhythmic response of the network in the form of trains associated with changes in the average values of the connections.  
During the stimulation, an important spiking synchronization occurs that causes, according to the relaxation 
process in $\eqref{inhomecha}$, a net increase of $\mathit{\phi_i}$  variables and their average value. 
This synchronization is maintained periodically (see also\blue{\cite{RodHil,Harris,Abbott2,Lindsey}}).

As shown in Fig.\,\blue{\subref*{NonUnif-c}}, this approach is part of a widely accepted rule of reinforcement of (maximum values of ) synaptic conductance for a given connection. This strengthening is linked to a causal relationship between spiking activity of  pre and postsynaptic neurons in this connection, see\blue{\cite{Abbott2}}. In this figure, the temporal behavior of the membrane activities $V_{pre}(t)$, $V_{post}(t)$ and strength  variables $W_{pre,post}(t)=\frac{J^{\mathit{E}}}{K}\Gamma(\phi_{pre}(t), \phi_{post}(t)) $ are represented in a superimposed manner for two  arbitrarily chosen cells  in the network, for a small number of trials.
Before the start of the stimulation, during the background activity, action potentials (APs) which are due to noise may be emitted in an uncoordinated way. The production of these APs does not cause a significant change in connection weights. During and after stimulation, and because of increased synchronization of pre- and postsynaptic activities, $\phi_{pre}(t), \phi_{post}(t)$ variables grow significantly and remain at a higher value.

On the other hand, for a network with inhibitory connections, the same mechanisms \eqref{inhomecha} with \eqref{OmegaFunc} and \eqref{Gamma1}, do not produce such 
sharp synchronization as in the previous case where the connections are only excitatory. Outside the period of stimulation, during which the  $\phi_i$ variables are growing, they decrease rapidly causing loss of rhythmicity, see Fig.\,\blue{\subref*{NonUnif-d}}.

\subsection {Two interacting Fitzhugh-Nagumo populations}
\label{susection:IntFiNapop}
\subsubsection {One cell connected to a network}
\label{subsubsec:1cell1Pop}
In section \bref{subsec:MFHH}, a method was presented for the mean field derivation of the dynamical activity of a set of large-scale interconnected neuronal populations. 
The general form of the interaction terms that allows one to deduce the system of integro-partial differential equations that must give the neuronal probability densities was determined in \eqref{gamma_2}. 
It may be useful to analyze this relationship in a simple case where some populations consist not
of a large number of cells but, in contrast, are composed of only a small number, and even consist of only one cell.
	The goal here is to show that a mean field derivation  may be obtained in the case of neural systems where only one cell is submitted to external stimulation, this cell being connected to a large network consisting of cells whose parameters may be different from those of the stimulated cell.
	In the notations of section \bref{subsec:MFHH}, $Z_1^{1 }$ is the dynamical variable of $\mathit{Cell}_1$  on which is applied an external stimulation, $\{Z_i^{2 }\}_{ i=1,2,\dots,K^2}$  denote the dynamical variables of cells which form a large-scale network $\mathscr{P}_2$, all of them being connected to $\mathit{Cell}_1$.
The stochastic differential systems which are attached to such neuronal assemblies are
\begin{align}
\frac{dZ_1^{1}}{dt}&=\mathcal{F}^{1}(Z_1^{1 })+\zeta _{1,t}^{1}+ \mathcal{ M}^{1 1}(Z_1^{1},Z_1^{1 })+\frac{1}{K^2}\sum _{j=1}^{K^{2 }}\mathcal{ M}^{1 2}(Z_1^{1},Z_j^{2 }) \\
\frac{dZ_i^{2 }}{dt}&=\mathcal{F}^{2}(Z_i^{2 })+\zeta _{i,t}^{2}+\mathcal{ M}^{2 1}(Z_i^{2},Z_1^{1 })+\frac{1}{K^2}\sum _{j=1}^{K^{2 }}\mathcal{ M}^{2 2}(Z_i^{2},Z_j^{2 }), \ \ i=1,2, \ldots, K^{2}.
\end{align}
In our approach, the $\mathscr{P}_2$  population is not submitted to an external stimulation, the term $\mathcal{ M}^{1 1}(Z_1^{1},Z_1^{1 })$ has the meaning of a self-interaction for $\mathit{Cell}_1$, $\mathcal{ M}^{2 1}(Z_i^{2},Z_1^{1 })$ refers to couplings of  $\mathit{Cell}_1$ to $\mathscr{P}_2$, ${\mathcal{ M}}^{1 2}(Z_1^{1},Z_j^{2 })$ represents the backward couplings from $\mathscr{P}_2$ to  $\mathit{Cell}_1$ and $\mathcal{ M}^{2 2}(Z_i^{2},Z_j^{2 })$ are internal interactions within $\mathscr{P}_2$. $\zeta _{1,t}^{1}$ and $\zeta _{i,t}^{2}$ are noise terms, the functions $\mathcal{F}^{j}, j=1,2$ are given in Section  \bref{subsec:DynSystOnePop}.

For such systems, the set variable of $\mathit{Cell}_1$ and  $\mathscr{P}_2$ are respectively $\widehat{n}{^{1}}(U)=\delta(Z_1^{1 }-U) $ and \\ $\widehat{n}{^{2 }}(U)=\frac 1{K^{2 }}\sum _{i=1}^{K^{2 }}\delta(Z_i^{2 }-U) $. The terms $\gamma^{\mu}_k,  k=1,2,3, \mu=1,2$ (see section \bref{subsec:MFHH}) were obtained in a general form. Let us examine more specifically the terms $\gamma^{\mu} _2$, as expressed in \eqref{gamma_2}, which have been obtained without any approximation. 
For $\mu=1$, the $\mathit{Cell}_1$ term is : 
\begin{equation}
\gamma^{1}_2=-\frac{\partial }{\partial U}\int_{R^{m+2}} \mathit{dU}\mathit{^\prime}\mathcal{M}^{1 1}(U,U^\prime)\langle\widehat{n}{^{1 }}   (U)\widehat{n}{^{1 }}    (U^\prime)\rangle_{p_t} -\frac{\partial }{\partial U}\int_{R^{m+2}} \mathit{dU}\mathit{^\prime}\mathcal{M}^{1 2}(U,U^\prime)\langle\widehat{n}{^{1 }}   (U)\widehat{n}{^{2}}    (U^\prime)\rangle_{p_t} 
\label{gamma1}
\end{equation} 
whereas for $\mu=2$, the  $\mathscr{P}_2$ term is :
\begin{equation}
\gamma^{2}_2=-\frac{\partial }{\partial U}\int_{R^{m+2}} \mathit{dU}\mathit{^\prime}\mathcal{M}^{2 1}(U,U^\prime)\langle\widehat{n}{^{2}}   (U)\widehat{n}{^{1 }}    (U^\prime)\rangle_{p_t} -\frac{\partial }{\partial U}\int_{R^{m+2}} \mathit{dU}\mathit{^\prime}\mathcal{M}^{2 2}(U,U^\prime)\langle\widehat{n}{^{2 }}   (U)\widehat{n}{^{2}}    (U^\prime)\rangle_{p_t}.
\label{gamma2}
\end{equation} 
Since $\mathscr{P}_1$ consists of only one cell, it is clear that in general, $\langle\widehat{n}{^{1 }}   (U)\widehat{n}{^{1 }}    (U^\prime)\rangle_{p_t}\neq\langle\widehat{n}{^{1 }}   (U)\rangle_{p_t}\langle\widehat{n}{^{1 }}(U^\prime)\rangle_{p_t}$.

However, this term only appears if a self-interaction is considered on $\mathit{Cell}_1$. Examination of the other terms in \eqref{gamma1},   \eqref{gamma2} leads to the same conclusions as those that were deduced in section \bref{subsec:MFHH}. Thus, if $\mathscr{P}_2$ is large-scale, admitting  fluctuations in $\widehat{n}{^{2 }}(U)$ of small amplitude, $\widehat{n}{^{2 }}(U)\simeq\langle\widehat{n}{^{2 }}(U)\rangle_{p_t}$ and $\langle\widehat{n}{^{1 }} (U)\widehat{n}{^{2 }}  (U^\prime)\rangle_{p_t}\simeq \langle\widehat{n}{^{1 }} (U)\rangle_{p_t}\langle\widehat{n}{^{2}}(U^\prime)\rangle_{p_t}$. Similarly, $\langle\widehat{n}{^{2}} (U)\widehat{n}{^{2 }}    (U^\prime)\rangle_{p_t}\simeq\langle\widehat{n}{^{2 }}(U)\rangle_{p_t}\langle\widehat{n}{^{2 }}(U^\prime)\rangle_{p_t}$.

Finally, when one considers a system consisting of a cell $\mathit{Cell}_1$ and a population $\mathscr{P}_2$, the  structure of mean field integro-partial differential equations is the same as that which has been  obtained previously except for the terms of self-interaction of $\mathit{Cell}_1$ which must be evaluated without approximation.

\begin{figure}[!h]
\centering
\subfloat[Response of $\mathit{Cell}_1$ to an external stimulation.]
{\includegraphics[width=0.45\textwidth]{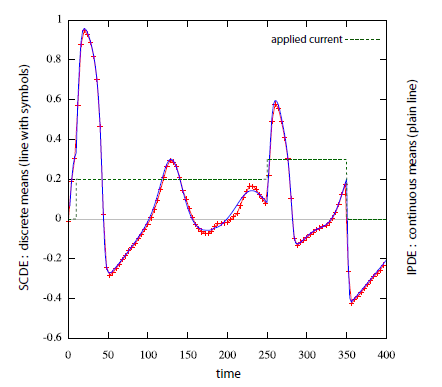} \label{ExciInh-a}} \\   
\subfloat[Response of $\mathscr{P}_2$ without coupling with $\mathit{Cell}_1$.]
{\includegraphics[width=0.45\textwidth]{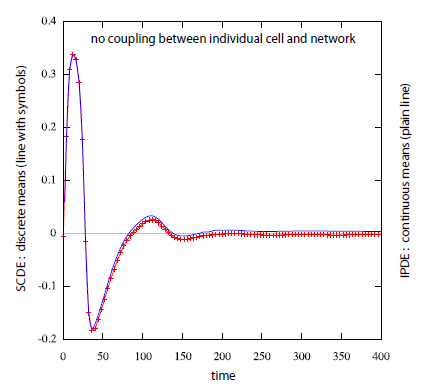} \label{ExciInh-b}}
\hfill
\subfloat[Response of $\mathscr{P}_2$ with coupling with $\mathit{Cell}_1$.]
{\includegraphics[width=0.45\textwidth]{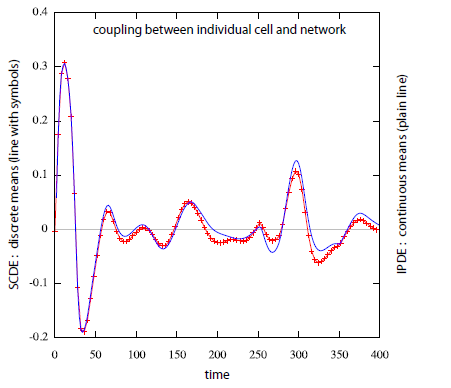} \label{ExciInh-c}}
\caption{Excitatory uniform connectivity in $\mathscr{P}_2$ with no external input. External input on $\mathscr{P}_1=\mathit{Cell}_1$. No backward connectivity from $\mathscr{P}_2$ to $\mathit{Cell}_1$. The stimulation of a single cell leads to a complex response of the $\mathscr{P}_2$ network which was analyzed by both methods.}
\label{1cell1Pop}
\end{figure}

In the numerical  illustration of "fill-in" which is shown in Fig.\,\blue{\ref{1cell1Pop}}, $\mathit{Cell}_1$ has no self-interaction and for clarity of the presentation, the backward coupling terms  are set to zero. The parameter values used in these simulations are  given in section \bref{sec:Num}. 

\subsubsection {Two large-scale excitatory and inhibitory connected populations}

\label{subsubsec:2popFN}
In this section, we consider two networks $\mathscr{P}_E$ and  $\mathscr{P}_I$ of neural populations with different parameters  
$k^{\alpha }$, $a^{\alpha }$, $b^{\alpha }$, $m^{\alpha }$,  $\alpha = E, I$. 
The connection matrix function   $\Psi=\left[\begin{matrix}M^{EE}&M^{EI}\\M^{IE}&M^{II}\end{matrix}\right]$, which
describes the connections inside each population and between the two populations, has its elements   $M^{\alpha \gamma }(V_i^{\alpha },X_i^{\alpha },\phi_i^{\alpha },V_j^{\gamma },X_j^{\gamma },\phi_j^{\gamma }), \ \alpha, \gamma =E, I$,  \ $i=1,2, \ldots, K_E, \ j=1,2, \ldots, K_I$ where  $K_E$ and  $K_I$ are the cell numbers of each population. As it has been shown in section \bref{sec:popdynsyst}, these matrix
elements may have a uniform character or a non uniform one which incorporates pre and postsynaptic aspects. In this section, we have considered the simplest form for these parameters,
the main objective here being to show how the integro-partial differential equations system  can help in the description and control of different coupled neural systems which are built of a great number
of cells. More precisely, all connections inside  $\mathscr{P}_E$ and  $\mathscr{P}_I$ are supposed uniformly distributed. 
As it has been considered in section \bref{subsubsec:ExcInhSynUnif}, the variables $\phi_i^{\alpha },   \ i=1,2,\ldots, K_{\alpha},\; \alpha=E,I$, assume constant values and synaptic currents between presynaptic cell $j$ and postsynaptic cell $i$ in  $\mathscr{P}_E$  (resp.  $\mathscr{P}_I$), are described by the term $ M^{EE}(V_i^E,X_i^E,V_j^E,X_j^E)=J^{\mathit{E}}(V_{S}^{\mathit{E}}-V_i^E)\, \sigma^{\mathit{E}}_{\Theta}(V_j^E) $
  (resp. $ M^{II}(V_i^I,X_i^I,V_j^I,X_j^I)=J^{\mathit{I}}(V_{S}^{\mathit{I}}-V_i^I)\, \sigma^{\mathit{I}}_{\Theta}(V_j^I))$. 
The connectivity between presynaptic cell $j$ in $\mathscr{P}_I$ (resp. $\mathscr{P}_E$) and postsynaptic cell $i$ in $\mathscr{P}_E$ (resp. $\mathscr{P}_I$) is described by $M^{EI}(V_i^E,X_i^E,V_j^I,X_j^I)=J^{EI}(V_S^{EI}-V_i^E)\, \sigma_{\Theta} ^{EI}(V_j^I)$ (resp. $M^{IE}(V_i^I,X_i^I,V_j^E,X_j^E)=J^{IE}(V_S^{IE}-V_i^I)\, \sigma_{\Theta} ^{IE}(V_j^E))$.

\begin{figure}[!ht] 
\captionsetup[subfigure]{justification=centering}
\centering             
\subfloat[Mean values of potential for $\mathscr{P}_E$. ]
{\put(55,120){\footnotesize No coupling between $\mathscr{P}_E$ and $\mathscr{P}_I$}  \includegraphics[width=0.47\textwidth]{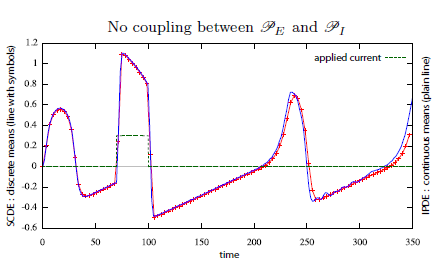} \label{2pops-a}}	\hfill
\subfloat[Mean values of potential for $\mathscr{P}_I$.]
{\put(55,120){\footnotesize No coupling between $\mathscr{P}_E$ and $\mathscr{P}_I$}  \includegraphics[width=0.47\textwidth]{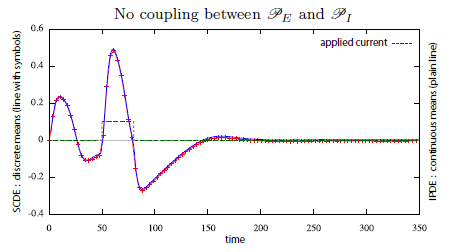}	\label{2pops-b}}	\\ [12pt]
\subfloat[Mean values of potential for $\mathscr{P}_E$.]
{\put(55,120){\footnotesize With coupling between $\mathscr{P}_E$ and $\mathscr{P}_I$}  \includegraphics[width=0.47\textwidth]{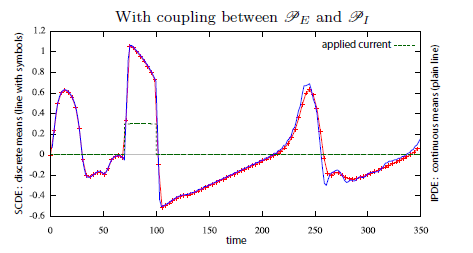}	\label{2pops-c}}	\hfill
\subfloat[Mean values of potential for $\mathscr{P}_I$.]
{\put(55,120){\footnotesize With coupling between $\mathscr{P}_E$ and $\mathscr{P}_I$}  \includegraphics[width=0.47\textwidth]{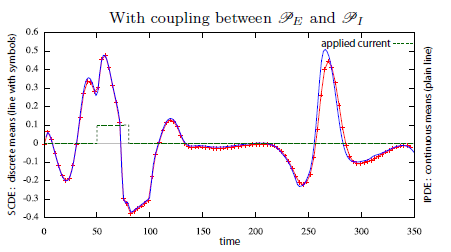}	\label{2pops-d}}	\\ [12pt]
\subfloat[Firing measures for $\mathscr{P}_E$, see \eqref{Fir1} and \eqref{Fir2}]
{\put(55,120){\footnotesize No coupling between $\mathscr{P}_E$ and $\mathscr{P}_I$}  \includegraphics[width=0.47\textwidth]{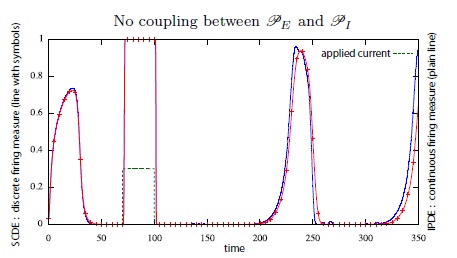}	\label{2pops-e}}	\hfill
\subfloat[Firing measures for $\mathscr{P}_I$, see \eqref{Fir1} and \eqref{Fir2}]
{\put(55,120){\footnotesize No coupling between $\mathscr{P}_E$ and $\mathscr{P}_I$}  \includegraphics[width=0.47\textwidth]{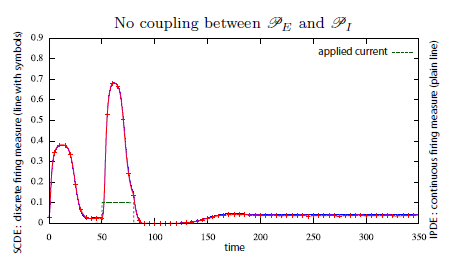}	\label{2pops-f}}	\\ [12pt]
\subfloat[Firing measures for $\mathscr{P}_E$, see \eqref{Fir1} and \eqref{Fir2}.]
{\put(55,120){\footnotesize With coupling between $\mathscr{P}_E$ and $\mathscr{P}_I$}  \includegraphics[width=0.47\textwidth]{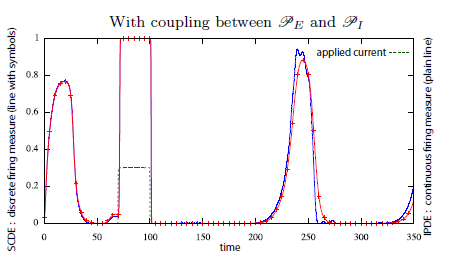}	\label{2pops-g}}	\hfill
\subfloat[Firing measures for $\mathscr{P}_I$, see \eqref{Fir1} and \eqref{Fir2}.]
{\put(55,120){\footnotesize With coupling between $\mathscr{P}_E$ and $\mathscr{P}_I$}  \includegraphics[width=0.47\textwidth]{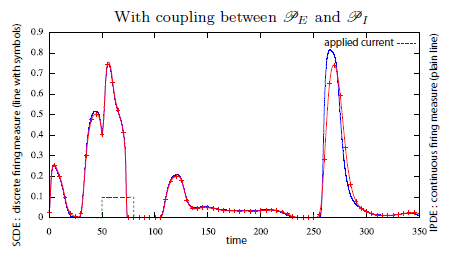}	\label{2pops-h}}\\ [10pt]
\caption{ Behavior of two coupled FN populations, one excitatory ($\mathscr{P}_E$) and one inhibitory ($\mathscr{P}_I$). }
\label{2pops} 
\end{figure}

As in previous sections, the numerical work is concentrated on a comparison between solutions of noisy dynamical systems, taking sufficiently many cells, and solutions which can be obtained from partial differential equations. 
The structure of the latter is made of a coupled system of equations for $n^E(V,X,t)$ and $n^I(V,X,t)$ which are the probability distributions for the excitatory and the inhibitory populations
\begin{align}
\frac{\mathit{\partial }}{\mathit{\partial }t}n^E(V,X,t)&=\frac{-\partial }{\partial V}\left\{(\mathit{F}^E(V,X)+I_{\mathit{ext}}^E(t))n^E(V,X,t)\right\}-\frac{\partial }{\partial X}\left\{\mathit{G}^E(V,X)n^E(V,X,t)\right\} \\
& -J^E\frac{\mathit{\partial }}{\mathit{\partial }V}\left\{(V_{S}^E-V)n^E(V,X,t)\right\}\int
_{R^2}\mathit{dV}\mathit{^\prime}\mathit{dX}\mathit{^\prime}\sigma _{\Theta}^E(V^\prime)n^E(V^\prime,X^\prime,t)
\nonumber  \\ 
& -J^{EI}\frac{\partial }{\partial V}\left\{(V_S^{EI}-V)n^E(V,X,t)\right\}\int
_{R^2}\mathit{dV}\mathit{^\prime}\mathit{dX}\mathit{^\prime}\sigma _{\Theta}^{EI}(V^\prime)n^I(V^\prime,X^\prime,t)+\frac 1
2(\beta _V^E)^2\frac{\partial ^2}{\partial V^2}n^E(V,X,t) \nonumber 
\end{align}
For $n^I(V,X,t)$ the equation is of the same type by making an exchange between the indices $I$ and $E$.
An example of numerical results is shown in Fig.\,\blue{\ref{2pops}}. 

\begin{itemize}[leftmargin=*]
\item Firstly, we have considered the time development of the probability densities in the case where the populations  $\mathscr{P}_E$  and $\mathscr{P}_I$ are uncoupled ($J^{EI}=J^{IE}=0$). 
An external input  $ I_{\mathit{ext}}^E(t)$ (resp.  $ I_{\mathit{ext}}^I(t))$ has been applied on $\mathscr{P}_E$ (resp. $\mathscr{P}_I$). 
The response of $\mathscr{P}_E$, which develops after applying the stimulus, appears as spiking activity in a temporal window (Fig.\,\blue{\subref*{2pops-a}})
where the firing probability is large - see Fig.\,\blue{\subref*{2pops-e}}. 
In $\mathscr{P}_I$, after application of the stimulus, the activity returns to an equilibrium (Fig.\,\blue{\subref*{2pops-b}}), in which the firing probability is small - see Fig.\,\blue{\subref*{2pops-f}}.

\item When the coupling between $\mathscr{P}_E$ and $\mathscr{P}_I$ is turned on ($J^{EI} \neq 0$, $ J^{IE} \neq 0$) and the same forms of stimuli are applied to the 2 populations, the response of $\mathscr{P}_E$, after application of the stimulus, does not change much as seen in Fig.\,\blue{\subref*{2pops-c}} and, similarly for the  firing probability  (Fig.\,\blue{\subref*{2pops-g}}). However, in the response of $\mathscr{P}_I$, we can see a significant difference from the uncoupled case. Indeed, the response in $\mathscr{P}_I$ is far from being a simple return to equilibrium. Significant spiking activity appears well after applying the stimulus (Fig.\,\blue{\subref*{2pops-d}}), which can be measured precisely in terms of the firing probability - see Fig.\,\blue{\subref*{2pops-h}}. Here also, we note good agreement between the solutions of $\mathrm{SCDE}$ and $\mathrm{IPDE}$.
\end{itemize}

\section{Numerics and parameter settings}
\label{sec:Num}
In this section, methods and parameters used in the various simulations of sections \bref{subsec:numres1pop} and \bref{susection:IntFiNapop} are presented. The numerical solutions of systems $\mathrm{SCDE}$ and $\mathrm{IPDE}$ were sought on the bounded domain  $\mathscr{D}~=~\Omega~\times~\Gamma~\times~\Phi$, such that $\Omega=[V_{min},V_{max}]=[-1.0,1.8]$, $\Gamma=[X_{min},X_{max}]=[-0.4,0.6]$, $\Phi=[\phi_{min},\phi_{max}]=[-1.0,3.0]$. The domain amplitudes have been chosen sufficiently large so that, in the numerical simulations, during the temporal evolution, the probability density and its derivatives are negligible outside  $\mathscr{D}$.

\begin{itemize}[leftmargin=*]
\item 
$\mathrm{\bf SCDE}$. An Euler method was used for the solution of systems $\mathrm{SCDE}$  with a time step $\delta {\rm t} = 0.01$. The required number of cells in each population was $K = 200$.  In each case $N = 500$ Monte Carlo simulations of the network equations were performed.
The same Gaussian distribution was chosen for the random initial conditions of these systems.  In the case of homogeneous and constant connections, the mean and dispersion parameters of this distribution for potential and recovery variables were set to $V_0=0$, $\sigma_V=0.15$, $X_0 =0$, $\sigma_X=0.15$. When dealing with the inhomogeneous case (section \bref{subsubsec:ExcSynNonUnif}), the corresponding Gaussian parameters for the  synaptic connections variables were set to $\phi_0=1.0$, $\sigma_{\phi}=0.01$.

For the determination of the discrete probability densities for the potential and synaptic connections variables (see \eqref{rhodisc} for $\xi^{\mu}(t)$), a partitioning of both $\Omega$ and $\Phi$ domains in N = 200 subintervals (bins) was performed. 
In our network model with excitatory synapses (section  \bref{subsubsec:ExcInhSynUnif}), detections of presynaptic activity causing the appearance of incoming synaptic currents in each cell was achieved by means of sigmoidal functions   (see \eqref{sigmoidV}) with $\beta=20$, $V^{\mathit{E}}_{\Theta}=0.5$. Other parameters were introduced to describe these currents,  $J^{\mathit{E}}=0.1$, $V_S^{\mathit{E}}=0.8$. In simulations involving 2 different populations (section  \bref{subsubsec:2popFN}), these parameters are $V_{S}^{E}=0.8$, $V_{S}^{I}=-0.2$, while internal coupling parameters for each population are $J^{E}=0.25$, $J^{I}=0.2$. Regarding the couplings between the two populations, one has $V_S^{EI}=1.0$, $V_S^{IE}=-1.0$, $J^{EI}=J^{IE}=0.2$.

Single populations of FN systems are expressed in terms of the following parameters $a=0.1$, $\gamma=0.2$, $k=1.0$, $b=0.015$, 
which are defined in \eqref{form1}, \eqref{form2} 
both for homogeneous (section  \bref{subsubsec:ExcInhSynUnif}) or inhomogeneous (section  \bref{subsubsec:ExcSynNonUnif}) connections. 
 In the case of two coupled homogeneous populations (section  \bref{subsubsec:2popFN}), these parameters have the following values: for $\mathscr{P}_E$, $a^E=0.1$, $\gamma^E=0.2$, $k^E=1.0$, $b^E=0.015$, and for $\mathscr{P}_I$, $a^I=0.15$, $\gamma^I=0.18$, $k^I=0.9$, $b^I=0.017$.

In section \bref{subsubsec:ExcSynNonUnif} we introduced a time-dependent inhomogeneous interneuronal connection model. For both excitatory and inhibitory networks, the dynamics of connection variables $\mathit{\phi_i},\, i=1,2,\dots, K =200$ depends on the parameters $\alpha_{\phi}=0.003$, $\beta_{\phi}=0.01$ and on a sigmoidal function $\sigma_{\phi}(X)=(1+e^{-\nu(X-X_{thresh})})^{-1}$, where $\nu=20.0$ and  $X_{S}=0.1$. In both cases $J=0.1$. In the excitatory case, $V_S^{\mathit{E}}=0.8$ while in the inhibitory case, $V_S^{\mathit{I}}=-0.2$.
 
 In section \bref{subsubsec:1cell1Pop} a cell ($\mathit{Cell}_1$)  is submitted to an external current $I(t)$. Internal parameters for $\mathit{Cell}_1$  are $a^1=0.1$, $\gamma^1=0.2$, $k^1=1.0$, $b^1=0.015$ .  $\mathit{Cell}_1$ is connected to a network $\mathscr{P}_2$ of cells whose parameters are $a^2=0.15$, $\gamma^2=0.18$, $k^2=0.9$, $b^2=0.017$.
$\mathscr{P}_2$  cells are not subjected  to external stimulation. The coupling is of forward type from $\mathit{Cell}_1$ to  $\mathscr{P}_2$ with $J_{12} =0.02$. The coupling from $\mathscr{P}_2$ to  $\mathit{Cell}_1$  is set equal to zero. Internal connections are of excitatory type, settings for these connections are identical to those considered in the previous sections.
All these systems undergo the action of white noise (see \eqref{noise}) for which the (diffusion)  parameter is set to $\beta_V=0.044 \; (\beta_V^2/2 \simeq 0.001)$.

\item 
$\mathrm{\bf IPDE}$. Solving the nonlinear $\mathrm{IPDE}$ of the (MVFP) type for $n(V,X,\phi,t)$ requires the knowledge of an initial condition $n(V,X,\phi,t=0)$.  We took a Gaussian with the same distribution parameters as the distribution of initial conditions used to solve the stochastic systems. In the general case of inhomogeneous connections, one has
\begin {equation}
n(V,X,\phi,t=0)=\frac{1}{(2\pi)^{3/2} \sigma _V \sigma _X \sigma _{\phi}}\exp\bigg(-\frac{(V-V_0)^2}{2\sigma_V^2}-\frac{(X-X_0)^2}{2\sigma _X^2}-\frac{(\phi-\phi_0)^2}{2\sigma _{\phi}^2}\bigg).
\end{equation}
Dirichlet boundary conditions were imposed on the bounded domain $\mathscr{D}$.  For computer simulation,  variable dimension $V \text{(resp. }X, \,\phi)$ of the 3-D  domain $\mathscr{D}$ has been discretized in $N_1$ (resp. $N_2, \,N_3$) points,  with $N_1= N_2= N_3=150.$

We have used an explicit midpoint Euler (order $2$) algorithm with time step $\delta {\rm t}=0.001$ which has been found sufficient to ensure stability of the numerical solution of the partial differential equation. During the simulation, regular monitoring of the normalization condition of the solution was made. 
\end{itemize}

\noindent Finally, the detection threshold $\theta$ of action potentials for the firing probabilities of both $\mathrm{SCDE}$ and $\mathrm{IPDE}$ (see \eqref{Fir1} and \eqref{Fir2}) was set at $\theta=0.8$. 
\medskip

Numerical simulations were performed using double precision floating point arithmetic on a Intel Xeon computer with 32 processors.

\section {Conclusion}
\label{sec:conclusion}
In recent work on the study of large-scale populations of neurons (see e.g.\blue{\cite{ToubFaug}}), mathematically rigorous methods from probability theory were developed. Mean field equations have been derived for a set of probability distributions governing the dynamical behavior of a set of noisy populations of neurons. These so called McKean-Vlasov-Fokker-Planck equations appear to bring a particularly interesting perspective in the field of neural networks.

We have developed a method which falls within the same framework of description of the macroscopic behavior of sets of conductance based neuronal populations in terms of the microscopic characteristics of the constituent nerve cells. This method uses some tools of kinetic theory of gases and plasmas\blue{\cite{Ichima}}. Here, the dynamics on the movements of the particles is replaced by the dynamics of the membrane depolarizations while the synaptic connections substitute for collisions between particles.

This approach of the neural behavior has been the subject of several studies in which cells are seen as coupled oscillators (see e.g.\blue{\cite{Buice2}}). This requires making simplifying assumptions (basic variables reduced to phases, weak coupling). The aim of this work is to propose a generalization of these methods to biologically plausible systems which are distributed in several interacting populations.

To achieve this goal, we have introduced stochastic set variables, as does Klimontovich \blue{\cite{Ichima}}, for cells belonging to all populations of the neural ensemble. Fokker Planck techniques thus lead to derivation of a hierarchical set of non closed non local integro-partial differential equations for the probability distributions of the neuronal dynamical variables.

However, these equations remain intractable and require, as in the kinetic theory of plasmas, suitable approximations for further analysis. Developments in statistical moments can be considered. We limited ourselves to a mean field approach that allows closure of the resulting hierarchy and is based on conventional estimates of the smallness of the fluctuations of set variables considered in the case of large neuronal populations.

Models of neurons that we have considered are of Hodgkin Huxley type, whose synaptic connection strengths are not necessarily constant over time, the proposed plasticity model being Hebbian. For such neuronal assemblies of interacting populations, we deduce McKean-Vlasov-Fokker-Planck type systems of equations similar to those obtained in\blue{\cite{ToubFaug}}.

Numerical integration of these equations has been made in the case of Fitzhugh-Nagumo neural networks and showed that statistical measurements obtained in this scheme are in good agreement with those obtained by direct simulations of finite size stochastic neural dynamic systems.

Kinetic theories have been developed in the framework of neural populations of oscillators, allowing analytical development for the study of statistical mechanics concepts, such as phase transitions and bifurcation analysis\blue{\cite{Kur1,Sakaguchi,Strogatz}}. The theory presented in this paper is a generalization for populations of conductance-based neurons.

While it seems impossible to consider such developments starting from the initial neural stochastic systems, it is hoped that the MVFP equations which have been obtained here, could bring new responses related to these statistical mechanics aspects.


\begin{thebibliography}{99}

\bibitem{Klimon} Yu. L. Klimontovich,  \textit{Kinetic Theory of Nonideal Gases and Nonideal Plasmas}. Pergamon Press (1982).

\bibitem{Ichima} S. Ichimaru,  \textit{Statistical Plasma Physics Volume I}. Addison-Wesley Publishing Company  (1992).

\bibitem{Nichol} D.R. Nicholson, \textit { Introduction to Plasma Theory}. Krieger, Malabar, Florida (1992).

\bibitem{Tuck1} H.C. Tuckwell, \textit{Introduction to Theoretical Neurobiology. Volume 1. Linear Cable Theory and Dendritic Structure - Volume 2. Nonlinear and Stochastic Theories}. Cambridge University Press (1988).

\bibitem{Tuck2} H.C. Tuckwell, \textit{Stochastic Processes in the Neurosciences}. SIAM (1989).

\bibitem{Tuck3} H.C. Tuckwell, \textit{Stochastic Partial Differential Equation Models in Neurobiology: linear and nonlinear models for spiking neurons}. Springer Lecture Notes in Mathematics, "Stochastic Biomathematical models", Chapter 6 (2013).

\bibitem{Tuck4} H.C. Tuckwell, \textit{Stochastic Modeling of Spreading Cortical Depression}. Springer Lecture Notes in Mathematics.,"Stochastic Biomathematical Models", Chapter 8 (2013).

\bibitem{Ermon} G.B. Ermentrout, D.H. Terman, \textit{Mathematical Foundations of Neuroscience, Interdisciplinary applied mathematics}. \textbf{35} Springer (2010).

\bibitem{RodTuck1} R. Rodriguez, H.C. Tuckwell, \textit{Statistical properties of stochastic nonlinear dynamical models of single spiking neurons and neural networks}. Physical Review E \textbf{54} (1996), 5585-5590.

\bibitem{Kur1} Y. Kuramoto, \textit{Chemical Oscillations, Waves, and Turbulence}. Springer-Verlag, Berlin (1984).

\bibitem{Kur2} Y. Kuramoto, \textit{Cooperative Dynamics of Oscillator Community A Study Based on Lattice of Rings}. Prog. Theor. Phys. Suppl.\textbf{79} (1984), 223-240.

\bibitem{KurNish} Y. Kuramoto, I. Nishikawa, \textit{in Cooperative Dynamics in Complex Physical Systems}, edited by H. Takayama Springer, New York (1988).

\bibitem{MirStrog} R.E. Mirollo, S.H. Strogatz, \textit{The spectrum of the locked state for the Kuramoto model of coupled oscillators}. Physica D \textbf{205} (2005), 249-266.

\bibitem{Sakaguchi} H. Sakaguchi, \textit{Cooperative Phenomena in Coupled Oscillator Systems under External Fields}. Progress of Theoretical Physics, \textbf{79} (1) (1988), 39-46.

\bibitem{Strogatz} S. H. Strogatz, \textit{From Kuramoto to Crawford: exploring the onset of synchronization in populations of coupled oscillators}. Physica D, \textbf{143} (2000), 1–20.

\bibitem{Abbott1} L.F. Abbott, C. van Vreeswijk, \textit{Asynchronous States in Networks of Pulse-Coupled Oscillators}. Physical Review E, \textbf{48}  (1993), 1483-1490. 

\bibitem{Daido} H. Daido, \textit{Onset of cooperative entrainment in limit-cycle oscillators with uniform all-to-all interactions : bifurcation of the order function}. Physica D \textbf{91}(1996), 24-66.

\bibitem{Buice1} E.J. Hildebrand, M.A. Buice and C.C. Chow, \textit{ Kinetic Theory of Coupled Oscillators}. Physical Review Letters \textbf{98} 054101 (2007).

\bibitem{McLaughlin} D. McLaughlin, R. Shapley,  M. Shelley, \textit{Large-scale modeling of the primary visual cortex: influence of cortical architecture upon neuronal response}. J. Physiol. Paris 97, (2003), 237-252.

\bibitem{Buice2} M.A. Buice  and C.C. Chow, \textit{Dynamic Finite Size Effects in Spiking Neural Networks}. PLOS Computational Biology (2013) pcbi.1002872 

\bibitem{Fasoli} D. Fasoli,  O. Faugeras, S. Panzeri, \textit{A formalism for evaluating analytically the cross-correlation structure of a firing-rate network model}. J. Math. Neurosci \textbf{5:6}, (2015), 1-53. 

\bibitem{Amit} D.J. Amit, N. Brunel, \textit{Dynamics of a recurrent network of spiking neurons before and following learning}. Network \textbf{8}, (1997) 373-404.  

\bibitem{Fasoli2} D. Fasoli, A. Cattani, S. Panzeri, \textit{The complexity of dynamics in small neural circuits}. arXiv preprint arXiv: 1506.08995 (2015).   

\bibitem{Tuck5} H.C. Tuckwell, \textit{Cortical network modeling: Analytical methods for firing rates and some properties of networks of LIF neurons}. J. Physiol. Paris 100, (2006) 88-99.

\bibitem{Tranch1} C. Ly, D. Tranchina, \textit{Critical Analysis of Dimension Reduction by a Moment Closure Method in a Population Density Approach to Neural Network Modeling}. Neural Computation \textbf{19}(2007), 2032-2092. 

\bibitem{NiCamp1} W. Nicola, S.A. Campbell, \textit{Bifurcations of large networks of two-dimensional integrate and fire neurons}. Journal of Computational Neuroscience \textbf{35} (1) (2013), 87-108. 

\bibitem{NiCamp2}W. Nicola, S.A. Campbell, \textit{Non-smooth bifurcations of mean field systems of two-dimensional integrate and fire neurons}. (2014), arXiv:1408.4767. 

\bibitem{Knight} B. W. Knight, \textit{Dynamics of Encoding in Neuron Populations: Some General Mathematical Features}. Neural Computation, \textbf{12} (2000), 473-518. 

\bibitem{Tranch2} F. Apfaltrer, C. Ly, D. Tranchina, \textit{Population density methods for stochastic neurons with realistic synaptic kinetics: Firing rate dynamics and fast computational methods}. Network: Computation in Neural Systems, \textbf{17} (2006), 373-418. 

\bibitem{Hansel} D. Hansel, G. Mato, \textit{Existence and stability of persistent states in large neural networks}. Physical Review Letters, \textbf{86}(18) (2001), 4175.

\bibitem{HH} A.L. Hodgkin, A.F. Huxley, \textit{A quantitative description of membrane current and its application to conduction and excitation in nerve}. J. Physiol.\textbf{ I I7} (I952), 500-544.

\bibitem{RodTuck2} R. Rodriguez, H.C. Tuckwell, \textit{Noisy spiking neurons and networks: useful approximations for firing probabilities and global behavior description}. Biosystems \textbf{48} (1998), 187-194. 

\bibitem{TucRod} H.C. Tuckwell, R. Rodriguez, \textit{Analytical and simulation results for stochastic Fitzhugh-Nagumo neurons and neural networks}. Journal of Computational Neurosciences \textbf{5} (1998), 91-113.   

\bibitem{ToubFaug} J. Baladron, D. Fasoli, O. Faugeras, J. Touboul, \textit{Mean-field description and propagation of chaos in networks of Hodgkin-Huxley and FitzHugh-Nagumo neurons}. J. Math. Neurosci  \textbf{2} (2012), 10.

\bibitem{Toub1} J. Touboul, \textit{Limits and Dynamics of Stochastic Neuronal Networks with Random Heterogeneous Delays}. J. Stat. Phys. \textbf{149} (2012), 569-597.

\bibitem{Toub2} S. Mischler, C. Quininao, J. Touboul, \textit{On a kinetic Fitzhugh Nagumo model of neuronal network}. arXiv:1503.00492v1 [math.AP] 2 Mar 2015.

\bibitem{FaugToubCessac} O. Faugeras, J. Touboul, B. Cessac, \textit{A constructive mean-field analysis of multi-population neural networks with random synaptic weights and stochastic inputs}. Front. Comp. Neurosci, \textbf{3}(1) (2009) 1-28.

\bibitem{DeMasi} A. De Masi, A. Galves, E. Löcherbach , E. Presutti,  \textit{ Hydrodynamic Limit for Interacting Neurons}. Journal of Statistical Physics, \textbf{158}(4) (2015) 866-902.

\bibitem{Song} S. Song, P. J. Sjöström, M. Reigl, S. Nelson, D. B. Chklovskii,  \textit{Highly Nonrandom Features of Synaptic Connectivity in Local Cortical Circuits}.  PLOS Biology, \textbf{3}(3) (2005) 0507-0519.

\bibitem{RodHil} P. Achard, S. Zanella, R. Rodriguez, G. Hilaire, \textit{Perinatal maturation of the respiratory rhythm generator in mammals: from experimental results to computational simulation}. Respiratory Physiology and Neurobiology \textbf{149} (2005), 17-27.

\bibitem{Harris} R.M. Harris-Warrick, \textit{General Principles of Rhythmogenesis in Central Pattern Networks}. Prog Brain Res. (2010) \textbf{187}, 213-222.

\bibitem{Abbott2} P. Dayan, L.F. Abbott, \textit{Theoretical Neuroscience: Computational and Mathematical Modeling of Neural Systems}. MIT Press (2005).

\bibitem{Lindsey} B.G. Lindsey, I.A. Rybak,  J. C. Smith, \textit{Computational Models and Emergent Properties of Respiratory Neural Networks}. Compr. Physiol.  \textbf{2}(3) (2012),1619-1670.

\bibitem{Brette} R. Brette, M. Rudolph, T. Carnevale, M. Hines, D.  Beeman, J.M. Bower, M. Diesmann, A. Morrison, P.H. Goodman, F.C. Harris Jr, M.Zirpe,   \textit{Simulation of networks of spiking neurons: a review of tools and strategies}. Journal of Computational Neuroscience,  \textbf{23}(3), (2007), 349-398.

\bibitem{Djurfeldt} M. Djurfeldt, Ö. Ekeberg, A. Lansner, \textit{Large-scale modeling-a tool for conquering the complexity of the brain}. Frontiers in Neuroinformatics \textbf{1} (2008),1-4. 
\end{thebibliography}
\end{document}